%% file: main.tex
\documentclass[prb,preprint]{revtex4-1} 
\include{preamble}

\include{macros}

\graphicspath{{figures/}}
\setcitestyle{authoryear,open={(},close={)}}
\begin{document}
\title{A Long-Range Ising Model of a \banet}

\author{Jeyashree Krishnan}
\email{krishnan@aices.rwth-aachen.de} 
\altaffiliation[permanent address: ]{MTZ, Pauwelstrasse 19, Level 3, D-52074, Aachen, Germany} 
\affiliation{Aachen Institute for advanced study in Computational Engineering Science(AICES) Graduate School, RWTH Aachen University, Germany}
\affiliation{Joint Research Center for Computational Biomedicine(JRC-Combine), RWTH Aachen University, Germany}

\author{Reza Torabi}
\email{rezatorabi@aut.ac.ir}
\affiliation{Department of Physics and Astronomy, University of Calgary, Calgary, Alberta, Canada}

\author{Edoardo Di Napoli}
\email{dinapoli@aices.rwth-aachen.de}
\affiliation{Aachen Institute for advanced study in Computational Engineering Science(AICES) Graduate School, RWTH Aachen University, Germany}
\affiliation{J\"{u}lich Supercomputing Center, Forschungszentrum J\"{u}lich, J\"{u}lich, Germany}

\author{Carsten Honerkamp}
\email{honerkamp@physik.rwth-aachen.de}
\affiliation{Institute for Theoretical Solid State Physics, RWTH Aachen University, Germany}
\affiliation{JARA-FIT, Aachen, Germany}

\author{Andreas Schuppert}
\email{schuppert@aices.rwth-aachen.de}
\affiliation{Aachen Institute for advanced study in Computational Engineering Science(AICES) Graduate School, RWTH Aachen University, Germany}
\affiliation{Joint Research Center for Computational Biomedicine(JRC-Combine), RWTH Aachen University, Germany}

\begin{abstract}

Networks that have power-law connectivity, commonly referred to as the scale-free networks, are an important class of complex networks\citep{albert_2002}. A heterogeneous mean-field approximation has been previously proposed for the Ising model of the \ba\ model of scale-free networks with classical spins on the nodes wherein it was shown that the critical temperature for such a system scales logarithmically with network size \citep{bianconi_2002, aleksiejuk_2002}. For finite sizes, there is no criticality for such a system and hence no \textit{true} phase transition in terms of singular behavior. Further, in the thermodynamic limit, the mean-field prediction of an infinite critical temperature for the system may exclude any true phase transition even then.

Nevertheless, with an eye on potential applications of the model on biological systems that are generally finite, one may still try to find approximations that describe the relevant observables quantitatively. Here we present an alternative, approximate formulation for the description of the Ising model of a \banet. Using the classical definition of magnetization, defined as the ensemble average of all spins in the network, we show that Ising models on a network can be well-approximated by a long-range interacting homogeneous Ising model wherein each node of the network couples to all other spins with a strength determined by the mean degree of the \banet. In such a \textit{effective \lrim} of a \banet, the critical temperature is directly proportional to the number of preferentially attached links added to grow the network. This dependence allows us to \enquote{control} the critical behavior of a \ba\ network by changing the model parameters.
The \lrim\ describes the magnetization of the majority of the sites with average or smaller than average degree better compared to the heterogeneous mean-field approximation. However, the heterogeneous mean-field approximation is better for predicting the onset at higher temperatures. 

Further, we show that the thermodynamic behavior of a scale-free network is between that of a lattice and that of a clique. The critical temperatures of lattice and clique form the lower and upper bounds, respectively, of the critical temperature of the \ba\ scale-free network. This approximation of an Ising model of a scale-free network to a \lrim\ allows us to make a direct comparison of a scale-free network to simple graphs such as lattices and cliques of the same size. The \lrim\ is the only homogeneous description of \ba\ networks that we know of.\\


\end{abstract}

\maketitle 
\clearpage
\section{Introduction}
\label{sec:intro}

The study of disorder and critical phenomena occurring in complex networks has been an area of extensive study in the last couple of decades \citep{dorogovstev_2007,rozenfeld_2008, strogatz_2001}. Owing to the non-trivial topology that is neither regular nor random, complex networks exhibit phase transitions that are markedly different from lattices or complete graphs \citep{ising_1925, barrat_2000, ferreira_2010, herrero_2002, gitterman_2000, lopes_2004}. Among these, many studies have extensively used the Ising paradigm to model criticality in real-world networks owing to its simplicity and broad applicability outside of statistical mechanics \citep{castellano_2009, stauffer_2006, aldana_2004, kumar_2000, krishnan_2019a, pastor_2015, pekalski_2001}.

An important class of complex networks is those that exhibit a power-law distribution and commonly referred to as the scale-free networks. \ba\ model is an algorithm that uses the preferential attachment mechanism to generate such scale-free networks. In this model, new nodes are added to existing nodes in the network proportional to the degree of the existing nodes until the overall network size is generated \citep{albert_2002}. 

A heterogeneous mean-field approximation has been proposed for the Ising model of a \banet\ with $\pm 1$ spins on the nodes showing that the critical temperature for such a system scales logarithmically with network size \citep{bianconi_2002, aleksiejuk_2002}. We refer to this mean-field approximation as heterogeneous or degree-weighted since the mean magnetization was calculated as the mean of spins weighted by their respective degrees. For finite sizes, there is no criticality for such a system and hence no \textit{true} phase transition in terms of singular behavior. Further, at the thermodynamic limit $(\nn \rightarrow \infty)$, this gives an infinite critical temperature for the system.

This is unlike the well-defined second-order phase transition that is exhibited by a regular lattice. Essentially this difference in critical behavior arises from the connectivity structure that is fed into the Hamiltonian of the Ising model -- short- and long-range connections in the case of a scale-free network (given by the adjacency matrix)\citep{albert_2002}; purely short-range connections in the case of a lattice (only nearest-neighbor coupling) \citep{ising_1925}. The heterogeneous mean-field approximation may over-represent the nodes with higher connections. Further, to our knowledge, it does not allow a direct comparison of Ising models of homogeneous (such as lattices or cliques) and heterogeneous (any complex network) structures. 

Here we present an alternative mean-field approximation of the Ising model of a \banet\ wherein we choose the classical definition of magnetization defined as the ensemble average of all spins in the network. Such a system exhibits up-down symmetry when $\mg = 0$, and breaking of symmetry when $\mg$ is non-zero. In this, we approximate the adjacency matrix of the \banet\ by an effective coupling constant, thereby transforming the network Hamiltonian to the Hamiltonian of a lattice. 

We show that the Ising model on a \banet\ can be well-approximated by a homogeneous \textit{effective \lrim} wherein each node of the network couples to all other spins with a strength determined to the mean degree of the \banet. This approximation allows us to make a direct comparison of a scale-free network to simple graphs such as lattices and cliques of the same size. With this, we show that the critical temperature of these network structures can be directly mapped from one to another, and the classical Ising model. The \lrim\ describes the magnetization of the majority of the sites with average or smaller than average degree better. Preliminary results of this work have been presented in the form of a talk and thesis \citep{krishnan_2019b, krishnan_2019c}.

The paper is organized as follows: in \sect\ \ref{sec:ordising} we present an approximation of the Ising model on a \banet\ by an effective homogeneous Ising model with long-range interactions and compare the full network with approximation numerically using Monte Carlo simulations; in \sect\ \ref{sec:error} we analyze the cost of this approximation; followed by \sect\ \ref{sec:comparison} where we use the \lrim\ to compare criticality in regular and scale-free structures. In \sect\ \ref{sec:mftcomp}, we compare the proposed approximation with the state-of-the-art and identify the temperature ranges at which each of these models fits best.
\clearpage
\section{Approximation of Ising model on a \banet}
\label{sec:ordising}

Consider the Hamiltonian of the Ising model with spins $\si = \pm 1$ on a ferromagnetically coupled \banet\ with $\nn$ nodes and $\mm$ preferentially attached links,

\begin{equation}
\hm = -\frac{1}{2} \sum_{i,j=1}^{\nn} \cm \si \sj-  \hh \sum_{i=1}^{\nn} \si \hspace{1cm} \cm = \cc \adj
\label{eq:hamiltonian}
\end{equation}

\noindent where $\hh$ is uniform magnetic field; $\cc$ is the coupling constant; $\adj$ is the adjacency matrix. Since the adjacency matrix $\adj$ is symmetric, the pre-factor $\frac{1}{2}$ is included to not count any pairs twice. 

\noindent The elements of the adjacency matrix $\adj$ are equal to one if there is a link between nodes $i$ and $j$ and zero otherwise. Unlike grid structures, different realizations of a \banet\ would result in a different adjacency matrix (however with similar connectivity). Therefore mean over multiple realizations of the adjacency matrix is a better estimate for \eq\ \ref{eq:hamiltonian}. The mean over many copies of the network then has a tensor structure \citep{bianconi_2002} (\cf\ \app\ \ref{sec:appdx} for summary of this method),

\begin{equation}
[\adj] = \pij = \frac{m}{2} \frac{1}{\sqrt{t_it_j}} = \frac{1}{2\mm\nn}\kk_{i}\kk_{j}
\label{eq:meanadj}
\end{equation}

\noindent where $\mm$ is the number of preferentially attached links to construct \banet; $\nn$ is the network size; and $\kk_{i}$ is the node degree $i$. Substituting \eq\ \ref{eq:meanadj} in \eq\ \ref{eq:hamiltonian} we have, 

\begin{equation}
\hm = - \frac{\cc}{4 \mm \nn} \sum_{i,j=1}^{\nn} \kk_{i} \kk_{j} \si \sj - \hh \sum_{i=1}^{\nn} \si
\label{eq:savg}
\end{equation}

\noindent We can see that the Hamiltonian has non-zero mutual couplings between all pairs $i,j$ of spins which however still vary in strength by the factors $\kk_{i}\kk_{j}$. Since $\kk$ follows a power law degree distribution it makes it challenging to evaluate \eq\ \ref{eq:savg}. Here we make an approximation by considering the first statistical moment of the degree distribution \ie\ $\kk = \mk$,

\begin{equation}
\sum_{i,j}^{\nn} \kk_{i} \kk_{j} \si \sj = \mk^{2} \sum_{i,j=1}^{\nn}\si\sj
\label{eq:approx}
\end{equation}

\noindent Setting all $\kk_{i}=\mk$ would mean that the degree of every node in the network $\kk_{i}$ is equal to the mean degree of the network $\mk$. If we use this approximation after \eq\ \ref{eq:meanadj}, it simply homogenizes the coupling constants between the spin pairs. This way we end up with a homogeneous model where all spins are coupled irrespective of their distance. A detailed analysis of the outcomes of this approximation is discussed in \sect\ \ref{sec:error}. Re-writing \eq\ \ref{eq:savg},

\begin{equation}
\hm \approx - \frac{J {\mk}^2}{4 \mm \nn} \sum_{i,j=1}\si \sj - \hh \sum_{i=1}^{\nn} \si
\label{eq:approxhm1}
\end{equation}

\noindent The mean degree, $\mk$ on a \banet\ can be approximated as (\cf\ \app\ \ref{sec:appdx}),

\begin{equation}
\mk \approx 2 \mm
\label{eq:meandeg}
\end{equation}

\noindent From \eqs\ \ref{eq:approxhm1} and \ref{eq:meandeg},

\begin{equation}
\hm \approx - \frac{\jeff}{2} \sum_{i,j=1}^{\nn} \si \sj - \hh \sum_{i=1}^{\nn} \si
\label{eq:approxhm2}
\end{equation}

\noindent where, 

\begin{equation}
\jeff = \frac{2\mm \cc}{\nn}
\label{eq:jeff}
\end{equation}

\noindent Comparing \eqs\ \ref{eq:hamiltonian} and \ref{eq:approxhm2}, we have approximated Ising model on a \banet\ by a long-range coupled homogeneous Ising model with an effective coupling constant acting between all pairs of spins, $\jeff = \frac{2\mm \cc}{\nn}$. This is unlike the classical Ising model of a two-dimensional lattice with nearest neighbor coupling only, i.e. has a finite coordination number. In our case, the coordination number is $\nn-1$. 

The mean-field Hamiltonian of the approximation is,

\begin{equation}
\hmf = \frac{1}{2} \jeff \nn^2 \mg^2 - (\jeff \nn \mg + \hh) \sum_{i=1}^{\nn} \si
\label{eq:hmf}
\end{equation}

\noindent with the thermodynamically configuration-averaged magnetization $\mg = \frac{1}{\nn} \langle \sum_{i=1}^{\nn} \si \rangle$. Using this mean-field Hamiltonian we can obtain the partition function which enable us to evaluate $\mg$ for uniform magnetic field, $\hh$ as,

\begin{equation}
\mg = \tanh(\bt \hh + \bt \jeff \nn \mg)
\label{eq:mag}
\end{equation}

\noindent The critical temperature $\tc$ obtained when $\hh = 0$ is, 
\begin{equation}
\tc = \frac{\jeff \nn}{\kb}
\label{eq:tc1}
\end{equation}

\noindent where $\kb$ is the Boltzmann constant. From \eqs\ \ref{eq:jeff} and \ref{eq:tc1} we have,

\begin{equation}
\tc = \frac{2 \mm \cc}{\kb}
\label{eq:tc2}
\end{equation}

From \eq\ \ref{eq:tc2} we note that $\tc$ scales linearly with coupling constant, $\cc$ and number of preferentially attached links, $\mm$. First, consider magnetization at $\hh = 0$. We can ask how the order parameter decreases as we tend towards the critical point. Just below $\tp = \tc$, $\mm$ is small, so we can Taylor expand \eq\ \ref{eq:mag} and use \eq\ \ref{eq:tc1} to obtain,

\begin{equation}
\mg \propto \pm \Bigg( \frac{\tc - \tp }{\tp}\Bigg)^{\frac{1}{2}}, \hspace{1cm} \tp < \tc
\label{eq:magtc}
\end{equation}

\noindent At $\tp = \tc$ and as $\hh \rightarrow 0$, from \eqs\ \ref{eq:mag} and \ref{eq:tc2} we have,

\begin{equation}
\mg \propto \hh^{\frac{1}{3}}
\label{eq:magh}
\end{equation}

\begin{figure}[!htb]
\includegraphics[scale=0.9]{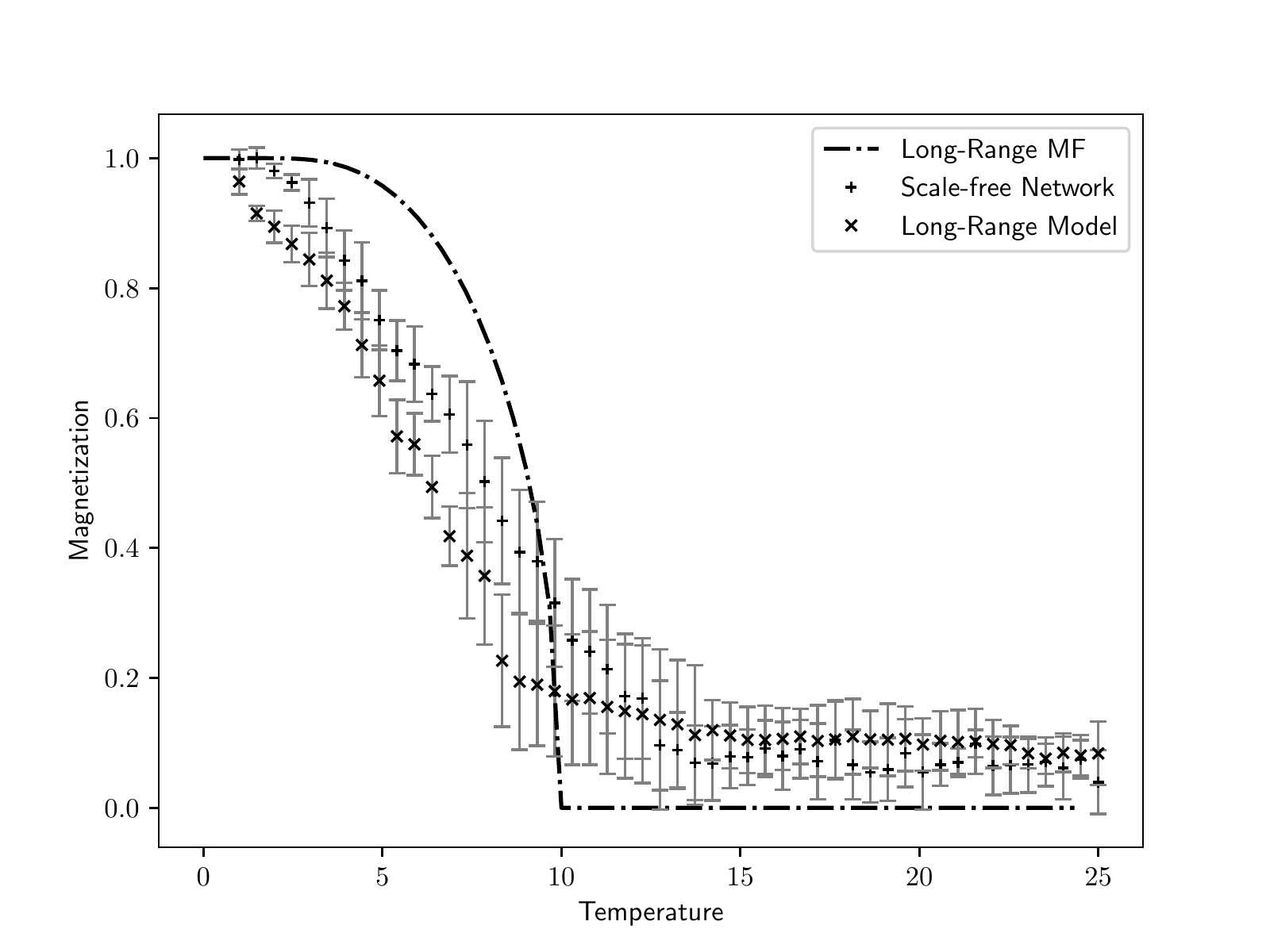}
\caption{ Comparison of the Ising model of \banet\ and \lrim\ of \banet: The plus markers show magnetization from the Monte Carlo simulations of the full ferromagnetically coupled \banet\ (given by \eq\ \ref{eq:hamiltonian}). The cross markers indicate the magnetization from the Monte Carlo simulations of the \lrim\ (given by \eq\ \ref{eq:approxhm2}). These data come from $n=20$ realizations of the \banet\ for the same choice of network parameters. The dotted-dashed line shows the mean-field solution of the \lrim\ (given by \eq\ \ref{eq:mag}). Simulation parameters: network size, $\nn = 5 \times 10^3$, preferential links added to grow the network, $\mm = 5$, coupling constant, $\cc = 1$ and magnetic field, $\hh = 0$.}
\label{fig:egcase}
\end{figure}

To compare the quality of this approximation, we performed Monte Carlo simulations of multiple realizations of the full \banet\ (\eq\ \ref{eq:hamiltonian} magnetization indicated as plus marks in \fig\ \ref{fig:egcase}) and the approximation \ie\ the \lrim\ (\eq\ \ref{eq:approxhm2} across multiple realizations indicated as cross marks in \fig\ \ref{fig:egcase}) using Metropolis local update algorithm \citep{metropolis_1953}. For a network of size $\nn = 5 \times 10^3$ and magnetic field $\hh = 0$, the system is equilibrated for $2 \times 10^4$ MC steps and thermodynamic variables sampled over $3 \times 10^4$ MC steps. Numerical $\tc$ is calculated when $\mg \approx 0.1$. 

As can be seen from \fig\ \ref{fig:egcase}, the magnetization values indicated by the Ising model on a \banet\ (indicated by plus markers) are well within the standard deviation of the \lrim\ (indicated by x markers) at very low and very high temperatures. At intermediate temperatures (and we will note later, at $\tp > \tc$), the effective ordinary Ising model underestimates magnetization. We also see that there is significant deviation of the numerical observations of the \lrim\ (shown by x markers) from the analytical solution (\eq\ \ref{eq:mag} shown by dotted dashed line). This is due to the small network size (order of thousands) that we consider here. The mean-field solution is expected to be exact for very large network sizes (when $\nn \rightarrow \infty$). 

At very low temperatures, the magnetization values indicated by the Ising model on a \banet\ are well within the standard deviation of the magnetization of the \lrim. At intermediate temperatures (and we will note later, at $\tp > \tc$), the effective ordinary Ising model underestimates magnetization. At very high temperatures, the magnetization values indicated by the \lrim\ agree well with the magnetization of the \banet. Further the results Monte Carlo simulations of the full \banet\ are in reasonable agreement with the mean-field approximation of \eq\ \ref{eq:mag} (\fig\ \ref{fig:egcase}). 

\begin{figure}[!htb]
\textbf{(A)} \includegraphics[scale = 0.35]{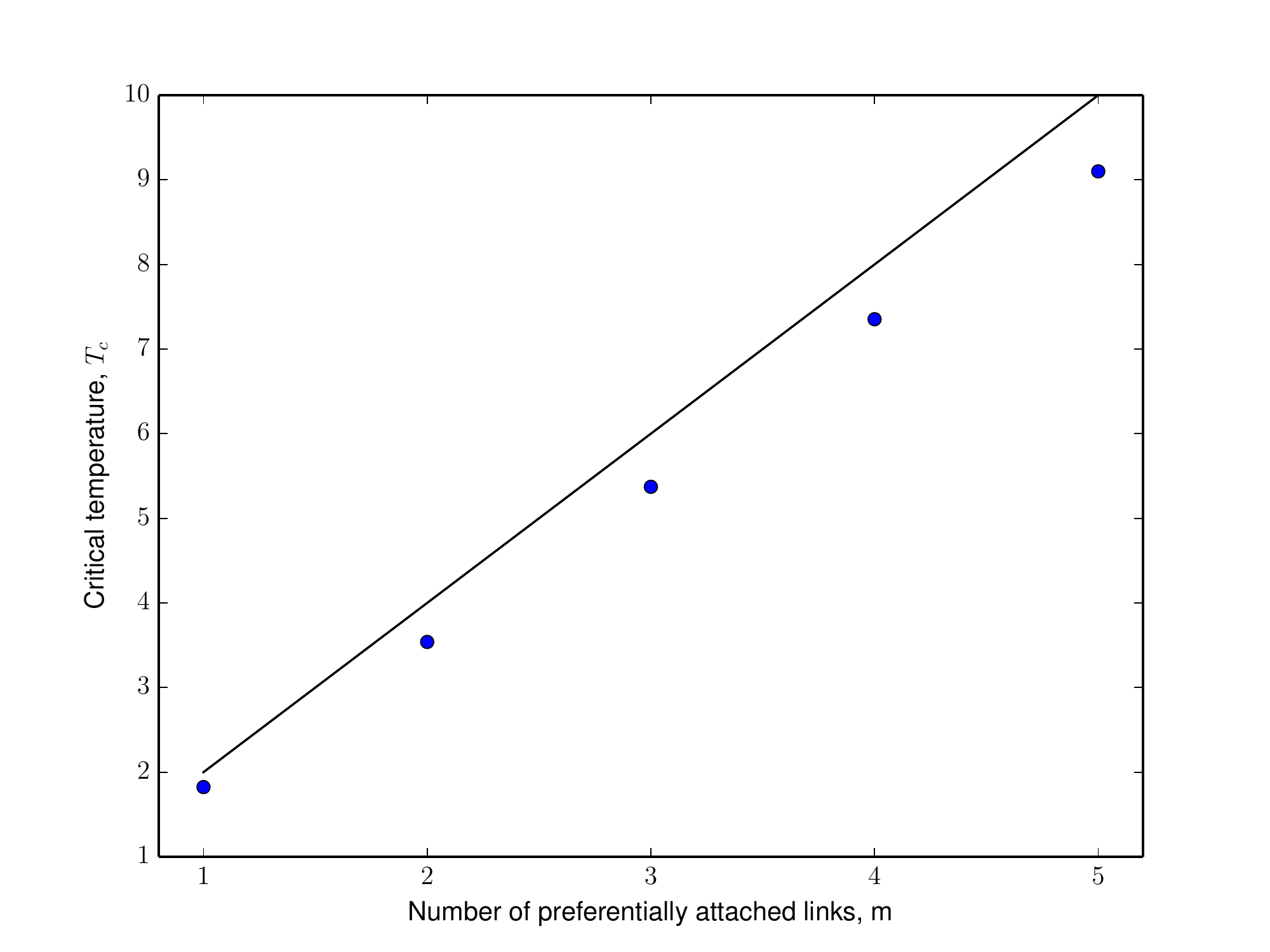}
\textbf{(B)} \includegraphics[scale = 0.35]{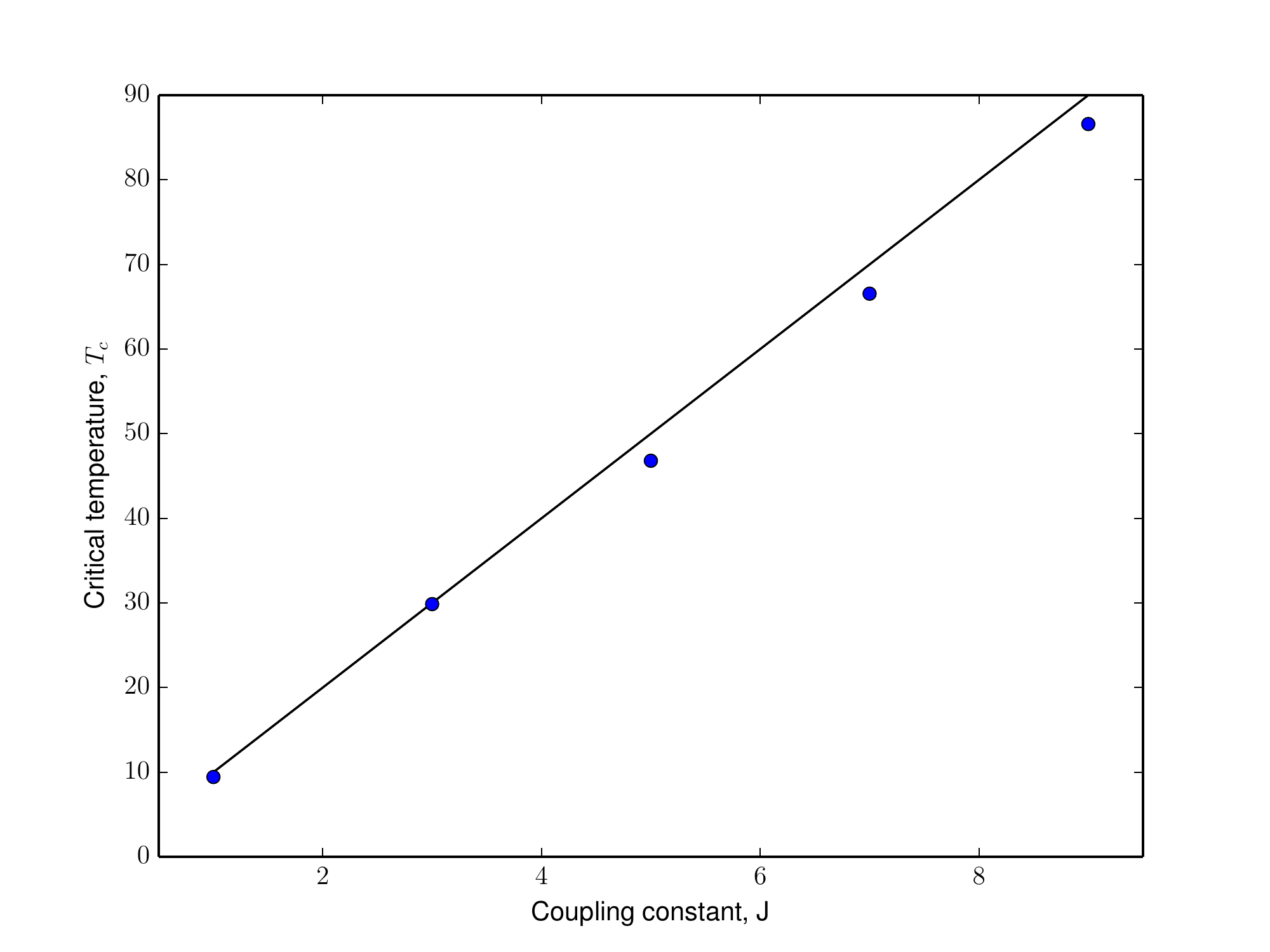}
\caption{Comparison of numerical and analytical results for scaling of critical temperature, $\tc$ with \textbf{(A)} preferentially attached links, $\mm$ and \textbf{(B)} coupling constant, $\cc$. The lines indicate the analytical approximation of $\tc$ from \eq\ \ref{eq:tc2} and the dots indicate the $\tc$ calculated from Monte Carlo simulations of the Ising model of the \banet\ from \eq\ \ref{eq:hamiltonian}. Numerical $\tc$ is calculated when $\mg \approx 0.1$.  }
\label{fig:scaling}
\end{figure}

As the coupling constant and the number of preferentially attached links increases, the system takes longer to reach a paramagnetic state. The numerical observations agree with the linear scaling of critical temperature with network parameters (\fig\ \ref{fig:scaling}). The slight deviation of numerical observations from expected scaling for high $\mm$ and $\cc$ could possibly be an effect of importance sampling in the Monte Carlo scheme. The model that comes out of such an approximation is interchangeably referred to as the \textit{\lrim} or \textit{$\mk$-clique model} in this paper.

\clearpage
\section{Approximation error}
\label{sec:error}

The approximation proposed in \sect\ \ref{sec:ordising} deviates from the expected trend, particularly when the system is at intermediate temperatures. Here we analyze the cost of the approximation of the Ising model on a \banet\ by a \lrim\ proposed in \sect\ \ref{sec:ordising}. In \ssec\ \ref{subsec:asympt} we do this by analyzing their asymptotic behavior and; in \ssec\ \ref{subsec:numerror} by numerically evaluating the neglected terms for different temperatures that lead to the deviation seen in \fig\ \ref{fig:egcase}.
\subsection{Asymptotic behavior of the \lrim}
\label{subsec:asympt}

Consider the approximation in \eq\ \ref{eq:approx} in \sect\ \ref{sec:ordising}. Re-writing the reduced Hamiltonian, 

\begin{equation}
\hm \approx - \frac{\cc}{4 \mm \nn} \sum_{i,j=1}^{\nn}\kk_{i}\kk_{j}\si\sj - \hh \sum_{i=1}^{\nn}\si \\
\label{eq:approxhm}
\end{equation}

To compare an Ising model on a \banet\ to a \lrim, we substitute,

\begin{equation}
\begin{split}
\kk_{i} &= \mk + \dk_{i}\\
\kk_{j} &= \mk + \dk_{j}\\
\end{split}
\label{eq:error}
\end{equation}

From \eqs\ \ref{eq:approxhm} and \ref{eq:error} we have,

\begin{equation}
\begin{split}
\hm &\approx - \frac{\cc}{4\mm\nn} \sum_{i,j=1}^{\nn} \Bigg( \mk^2 + 2 \mk \dk_{i} + \dk_{i}\dk_{j} \Bigg) \si\sj - \hh \sum_{i=1}^{\nn}\si\\
&\approx \Bigg( \frac{-\cc \mk^2}{4\mm\nn} \sum_{i,j=1}^{\nn} \si\sj - \hh \sum_{i=1}^{\nn} \si \Bigg) - \frac{\cc\mk}{2\mm\nn} \sum_{i,j=1}^{\nn} \dk_{i}\si\sj - \frac{\cc}{4\mm\nn}\sum_{i=1}^{\nn}\dk_{i}\dk_{j}\si\sj\\
\end{split}
\label{eq:hmerr}
\end{equation}

The first terms in brackets are the Hamiltonian for a \lrim\ with an effective coupling constant $\jeff = \frac{\cc \mk^2}{2\mm\nn}$. The additional terms are first and second-order error terms, respectively arising due to the approximation. We know from the simulation of an Ising model on \banet\ that the system is ordered at $\tp \rightarrow 0$, and all the spins are either $+1$ or $-1$. As $\tp \rightarrow \infty$ the spins are randomly distributed and the system is disordered (\fig\ \ref{fig:egcase}). Consider their asymptotic behavior:
\subsubsection{Ordered phase: $\tp \rightarrow 0$}
\label{subsubsec:ordered}

In this limit the contribution of the error terms are zero:

\begin{equation}
\begin{split}
- \frac{\cc \mk}{2\mm\nn} \sum_{i=1}^{\nn} \dk_{i}\si\sj &= - \frac{\cc \mk}{2 \mm \nn} \sum_{i=1}^{\nn}\sum_{i=1}^{\nn}\dk_{i}\\
&= 0\\
\end{split}
\label{eq:t0}
\end{equation}

\begin{equation}
\begin{split}
\frac{-\cc}{4\mm\nn} \sum_{i,j}^{\nn}\dk_{i}\dk_{j}\si\sj & =\frac{-\cc}{4\mm\nn}\sum_{i=1}^{\nn}\dk_i\si  \sum_{j=1}^{\nn}\dk_j\sj\\
&= 0\\
\end{split}
\end{equation}

\noindent since $\sum_{i=1}^{\nn} \dk_{i} = \sum_{j=1}^{\nn} \dk_{j} = 0$ (\cf\ \app\ \ref{sec:appdx}). Therefore the contribution of the two extra terms are zero at $\tp \rightarrow 0$. This means that at $\tp \rightarrow 0$, we can map the system to a \lrim\ with an effective coupling constant $\jeff = \frac{\cc \mk^2}{2\mm\nn}$. This encourages us to use a \lrim\ with an effective coupling constant and we can estimate $\tc$ as, $\frac{2\mm\cc}{\kb}$. 

In this temperature range (at $\tp \approx 1$), we see that although our estimation may not be very accurate it predicts a linear trend of variation of $\tc$ with respect to $\mm$ and $\cc$ (\cf\ \figs\  \ref{fig:egcase} and \ref{fig:scaling}). This is also reflected in the numerical evaluation of the approximation error at low temperatures (\cf\ $\tp \approx 1$ in \tab\ \ref{tab:error}).
\subsubsection{Disordered phase: $\tp \rightarrow \infty$}
\label{subsubsec:disordered}

In this limit the contribution of the second term is again zero:

\begin{equation}
\begin{split}
\frac{-\cc}{2\mm\nn} \sum_{i,j}^{\nn} \dk_{i}\si\sj &= 
\frac{-\cc}{2\mm\nn} \Bigg( \sum_{j=1}^{\nn}\sj\Bigg) \sum_{i,j}^{\nn}\dk_{i}\si\\
&= 0\\
\end{split}
\label{eq:tn}
\end{equation}

Because $\sum_{j=1}^{\nn}\sj = 0$ in this case (when all spins are random). However, the contribution of the third term is not zero in this limit:

\begin{equation}
\begin{split}
\frac{-\cc}{4\mm\nn} \sum_{i,j}^{\nn}\dk_{i}\dk_{j}\si\sj &= -\frac{\cc}{4\mm\nn} \Bigg( \sum_{i=1}^{\nn} \dk_{i}\si\Bigg)
\Bigg( \sum_{j=1}^{\nn}\dk_{j}\sj \Bigg)\\
&= - \frac{\cc}{4\mm\nn} \Bigg( \sum_{j=1}^{\nn}\dk_{j}\sj \Bigg)^2\\
&< 0\\
\end{split}
\end{equation}

This causes the deviation in magnetization at $0 < \tp < \tc$ between the Ising model of a \banet\ and the proposed approximation as can be seen in \fig\ \ref{fig:egcase}. This is also reflected in the numerical evaluation of the approximation error at high temperatures (\cf\ $\tp  < \tc$, $\tp \approx \tc$ and $\tp \approx 2\tc$ in \tab\ \ref{tab:error}).
\subsection{Numerical Evaluation of the Approximation Error}
\label{subsec:numerror}

Consider the degree distribution term in \eq\ \ref{eq:approxhm1},

\begin{equation}
\sum_{i,j}^{\nn}\kk_i\kk_j\si\sj
\label{eq:degdist}
\end{equation}

which is approximated by the mean degree. Re-writing \eq\ \ref{eq:error} we have,

\begin{equation}
\begin{split}
\kk_{i} &= \mk + \dk_{i}\\
\kk_{j} &= \mk + \dk_{j}\\
\end{split}
\end{equation}

Plugging \eq\ \ref{eq:error} in \eq\ \ref{eq:approxhm1},

\begin{equation}
\underbrace{\sum_{i,j}^{\nn} \kk_{i}\kk_{j}\si \sj}_{\mathrm{LHS}} =\underbrace{\mk^2 \sum_{i,j}^{\nn} \si\sj}_{t1} + 2\underbrace{\mk^2 \sum_{i,j}^{\nn} \dk_{i} \si\sj}_{t2} + \underbrace{\sum_{i,j}^{\nn} \dk_{i} \dk_{j} \si\sj}_{t3}
\label{eq:error1}
\end{equation}

The approximation presented in \sect\ \ref{sec:ordising} truncates the contribution of the degree distribution in \eq\ \ref{eq:degdist} to $t_1$. This approximation is hence a zero-order approximation in $\dk$. A re-scaling of the square of the mean degree term in $t_1$, $\mk^2$ allows us to map the Ising model on a \banet\ to the \lrim. $t_2$ includes fluctuation terms arising from the deviation of node degree from mean degree. For nodes with a very high degree, depending on the temperature, these terms may be high. $t_3$ is the second-order contribution of these fluctuation terms, and hence its contribution can be very high depending on the temperature.

\tab\ \ref{tab:error} summarize the contribution of each term at different temperature ranges evaluated numerically. The sum of the contribution of all terms (LHS) is very high at low temperatures ($\tp \approx  1$) for all choice of network parameters, given that all spins take $+1$ or $-1$ values for this temperature range. This decreases by order of magnitude at temperature ranges between low and critical temperature ($\tp \approx \frac{\tc}{2}$) where the spin configuration can be either $+1$ or $-1$. At critical temperature ($\tp \approx \tc$), LHS is another order of magnitude lower since the effect of most spins cancels each other.

Let us now consider observations for \banet\ of $\nn = 5 \times 10^3$ (\tab\ \ref{tab:error} (A)(B)(C)). At $\tp \approx \tp_0$ (ordered phase in \sssec\ \ref{subsubsec:ordered}), $t_1$ contributes most. In other words, the network can be well-approximated by the $\mk-$clique model at this temperature as can be verified from numerical simulations in \fig\ \ref{fig:egcase} (\cf\ limit cases \ssec\ \ref{subsec:asympt}). At temperatures close to critical temperature $0 < \tp < \tc$ (disordered phase in \sssec\ \ref{subsubsec:disordered}) the deviation from mean degree owing to high degree nodes causes $t_4$ to explode causing the deviation observed in numerical simulations in \fig\ \ref{fig:egcase}. At $\tp \approx \tc$ and $2 \tc$, close to paramagnetic state, $t_1$ contribution is high in most cases (not in (C)). The net contribution (LHS) may be negative since most spins may take values $-1$ (\cf\ limit cases \ssec\ \ref{subsec:asympt}). \tab\ \ref{tab:error} (D) summarizes approximation error for a different choice of network size, $\nn = 10^3$.

Overall, at ordered phases (very low temperatures or temperatures higher than critical temperatures), the proposed approximation is reasonable. At intermediate temperatures, there is a significant deviation between the approximation and the \banet\ owing to the dominance of the fluctuation term $t_3$. Overall, the Monte Carlo simulations of the \lrim\ proposed here (\eq\ \ref{eq:approxhm2}) is a modest approximation of the full \banet\ (\eq\ \ref{eq:hamiltonian}).

\begin{table}[!htb]
\label{tab:error}
  \textbf{(A)} \begin{tabular}{|c | c | c | c| c | c |}
  \hline
    Temperature, $\tp$ & $t_1$ & $t_2$ & $t_3$ & $t_4$ & LHS \\ \hline 
    $T \approx T_0$ & $499000.5$ & $9.9 \times 10^{-9}$ & $9.9 \times 10^{-9}$ & $-87.88$ & $498912.62 $\\ \hline 
    $T_0 < T < T_c$ & $612.77$ & $275.93$ & $275.93$ & $85392.04$ & $86556.67$\\ \hline        
    $T \approx T_c$ & $660.67$ & $-364.27$ & $-364.27$ & $-35.07$ & $103.34$ \\ \hline   
    $T \approx 2 T_c$ & $25.74$ & $7.1 \times 10^{-9}$ & $7.1 \times 10^{-9}$ & $-29.54$ & $3.8$ \\ \hline  
  \end{tabular}\\
  \vspace{1cm}
  \textbf{(B)} \begin{tabular}{|c | c | c | c| c | c |}
  \hline
    Temperature, $\tp$ & $t_1$ & $t_2$ & $t_3$ & $t_4$ & LHS \\ \hline 
    $T \approx T_0$ & $179784.06$ & $1.73 \times 10^{-10}$ & $1.73 \times 10^{-10}$ & $-619.32$ & $179164.74$\\ \hline 
    $T_0 < T < T_c$ & $-172.59$ & $-418.05$ & $-418.05$ & $30406.41$ & $29397.72$\\ \hline        
    $T \approx T_c$ & $432.11$ & $-615.45$ &  $-615.45$ & $97.29$ & $-701.5$ \\ \hline   
    $T \approx 2 T_c$ & $-12.94$ & $1.73 \times 10^{-10}$ & $1.73 \times 10^{-10}$ & $-9.64$ & $-22.58$ \\    
    \hline 
  \end{tabular}\\
  \vspace{1cm}
  \textbf{(C)} 
  \begin{tabular}{|c | c | c | c| c | c |}
  \hline
    Temperature, $\tp$ & $t_1$ & $t_2$ & $t_3$ & $t_4$ & LHS \\ \hline 
    $T \approx T_0$ & $19992.0008$ & $-7.43 \times 10^{-9}$ & $-7.43 \times 10^{-9}$ & $211.23$ & $20203.23$\\ \hline 
    $T_0 < T < T_c$ & $-24.55$ & $0.03$ & $0.03$ & $3631.96$ & $3607.47$\\ \hline
    $T \approx T_c$ & $9.67$ & $-44.98$ & $-44.98$ & $10.01$ & $-70.28$ \\  \hline 
    $T \approx 2 T_c$ & $2.07$ & $2.5 \times 10^{-10}$ & $2.5 \times 10^{-10}$ & $-3.53$ & $-1.46$ \\    
    \hline 
  \end{tabular}\\
  \vspace{1cm}
  \textbf{(D)}
  \begin{tabular}{|c | c | c | c| c | c |}
  \hline
    Temperature, $T$ & $t_1$ & $t_2$ & $t_3$ & $t_4$ & LHS \\ \hline 
    $T \approx T_0$ & $99002.5$ & $4.7 \times 10^{-10}$ & $4.7 \times 10^{-10}$ & $-436.90$ & $98565.6$\\ \hline 
    $T_0 < T < T_c$ & $310.86$ & $77.67$ & $77.67$ & $16339.75$ & $ 16805.95$\\ \hline        
    $T \approx T_c$ & $362.34$ & $-691.69$ & $-691.69$ & $4.78$ & $-1016.26$ \\  \hline  
    $T \approx 2 T_c$ & $10.89$ & $5.12 \times 10^{-10}$ & $5.12 \times 10^{-10}$ & $-19.25$ & $-8.36$ \\  
    \hline 
  \end{tabular}
  \caption[Numerical evaluation of approximation error]{ Numerical evaluation of approximation terms in \eq\ \ref{eq:error1} averaged over $100$ realizations of \banet (up to third decimal). Terms $t_1$ to $t_4$ evaluated at (a) $\tp \approx 1$ (top rows), (b) at $1 < \tp < \tc$,  (c) at analytical $\tc$ and (d) $\tp \approx 2\tc$. $\tc$ is calculated from \eq\ \ref{eq:tc2} for network parameters: \textbf{(A)} $\nn = 5000, \mm = 5$, \textbf{(B)} $\nn = 5000, \mm = 3$, \textbf{(C)} $\nn = 5000, \mm = 1$,  \textbf{(D)} $\nn = 1000, \mm = 5$. LHS is the sum of all terms on the left hand side of \eq\ \ref{eq:error1}.}
\end{table}
\clearpage
\section{Ising Model of Lattice, Scale-free network and Clique and their Relationships}
\label{sec:comparison}

The approximation of a scale-free network such as the \banet\ to a lattice-like Ising system now allows us to make a direct comparison of complex networks to simpler structures such as lattices and cliques. Let us consider mean-field approximations of the Ising model of the three topologies:
\subsubsection{Classical Ising Model of a Lattice}
\label{subsubsec:lattice}
This is the well-established two-dimensional Ising model of a quadratic lattice where only short-range nearest-neighbor interactions are allowed, as illustrated in \fig\ \ref{fig:struc} \textbf{(A)}. The number of interacting spins on a spin, $\zz$, is equal to the number of nearest neighbors, $\zz = 4$ and critical temperature $\tc$ is given by \citep{ising_1925},

\begin{equation}
\tc^{\mathrm{lattice}} = 2.27\cc
\label{eq:tclat}
\end{equation}

\subsubsection{Ising Model of a Clique}
\label{subsubsec:clique}

In a $\kk-$clique, all nodes interact with each other since it is a complete graph (as illustrated in \fig\ \ref{fig:struc} \textbf{(C)}).The number of interacting spins on clique is $(\nn-1)$, where $\nn$ is the total number of spins and the critical temperature $\tc$,

\begin{equation}
\tc^{\mathrm{clique}} = \frac{(\nn-1)\cc}{\kb} \approx \frac{\nn \cc}{\kb}
\label{eq:tcclq}
\end{equation}

\noindent since $\nn >> 1$.
\subsubsection{Ising Model of a Scale-Free Network}
\label{subsubsec:sfnet}

\begin{figure}[!htb]
    \centering
    \includegraphics[scale=0.4]{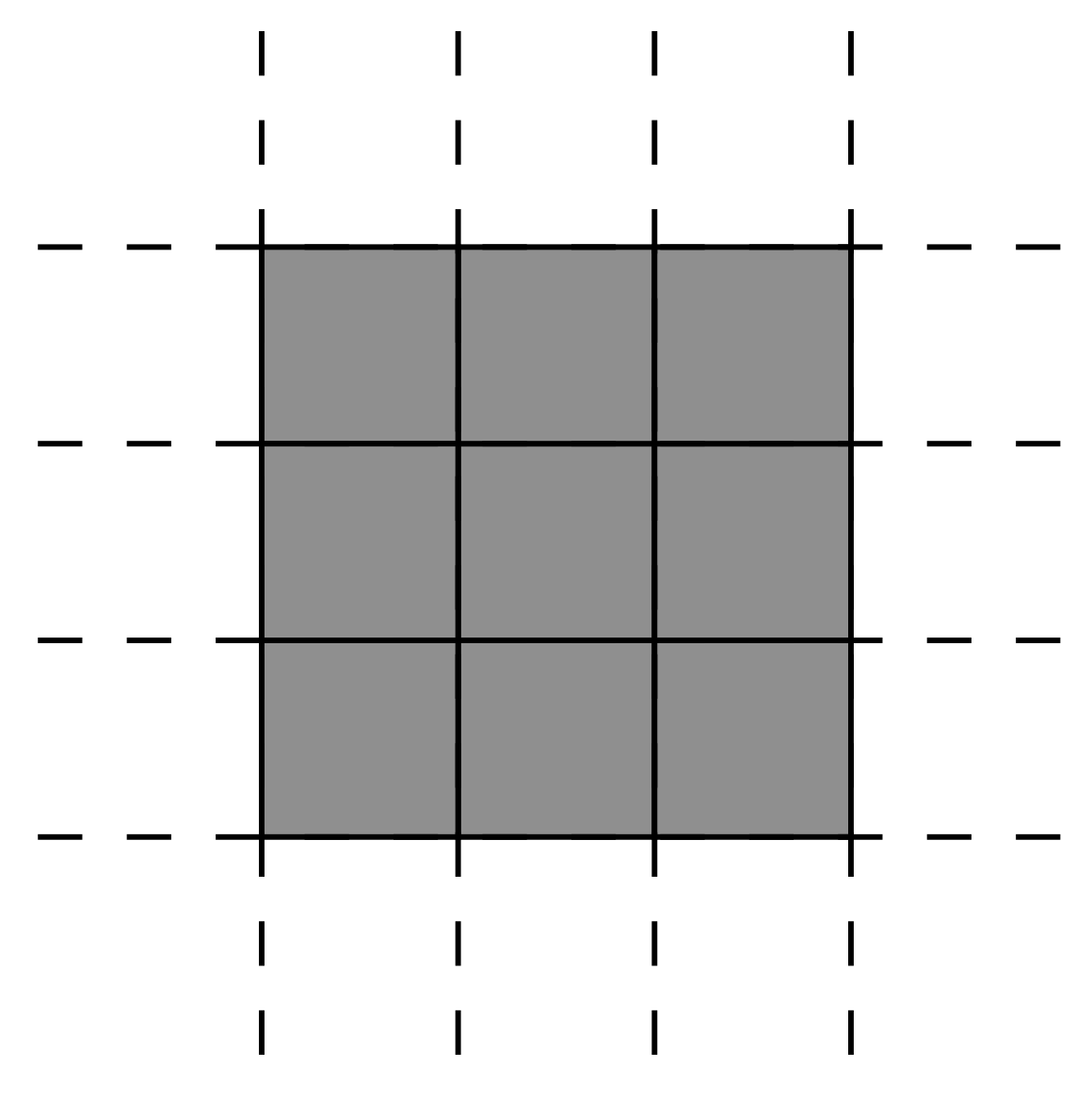}\\
    \includegraphics[scale=0.4]{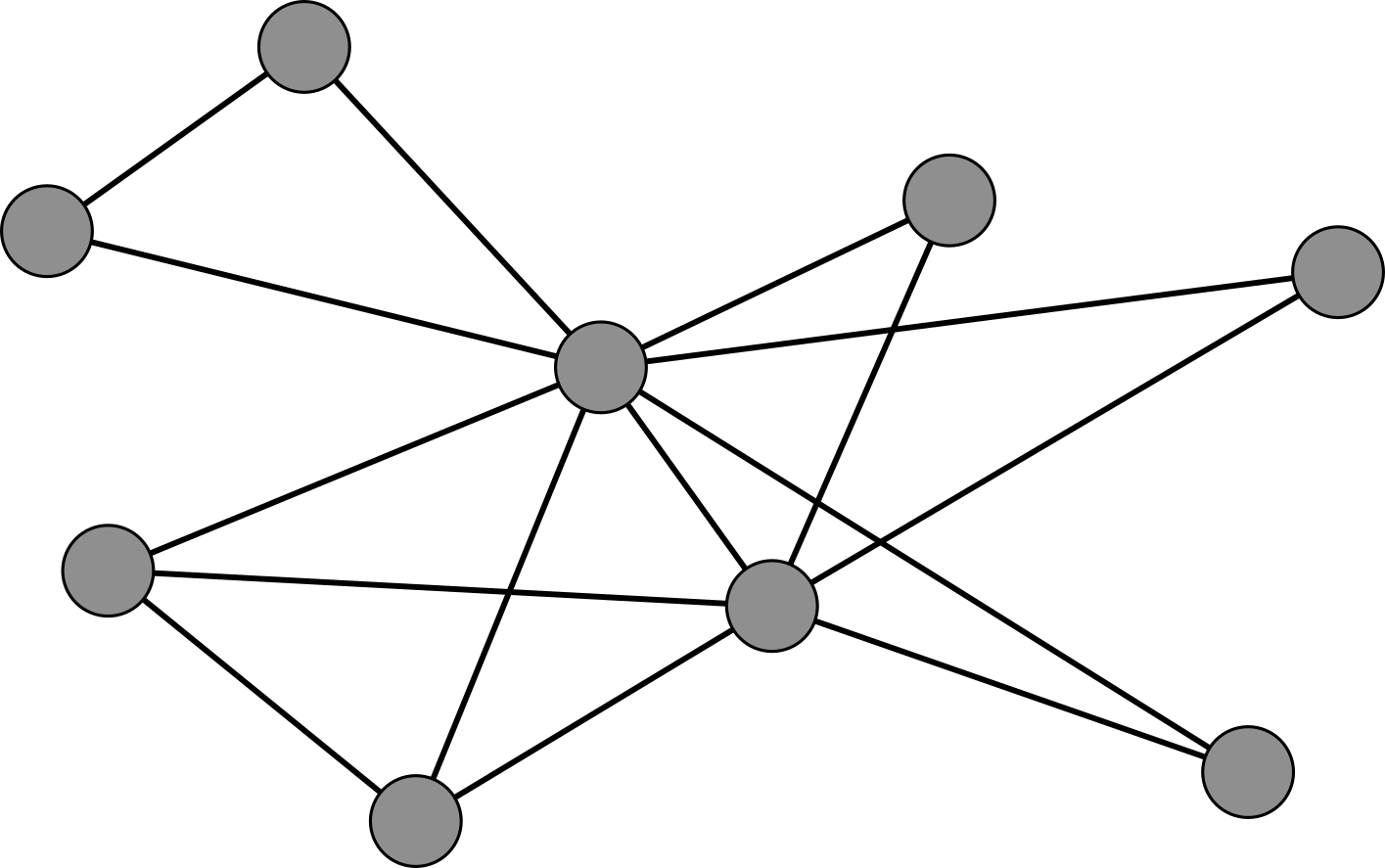}\\
    \includegraphics[scale=0.4]{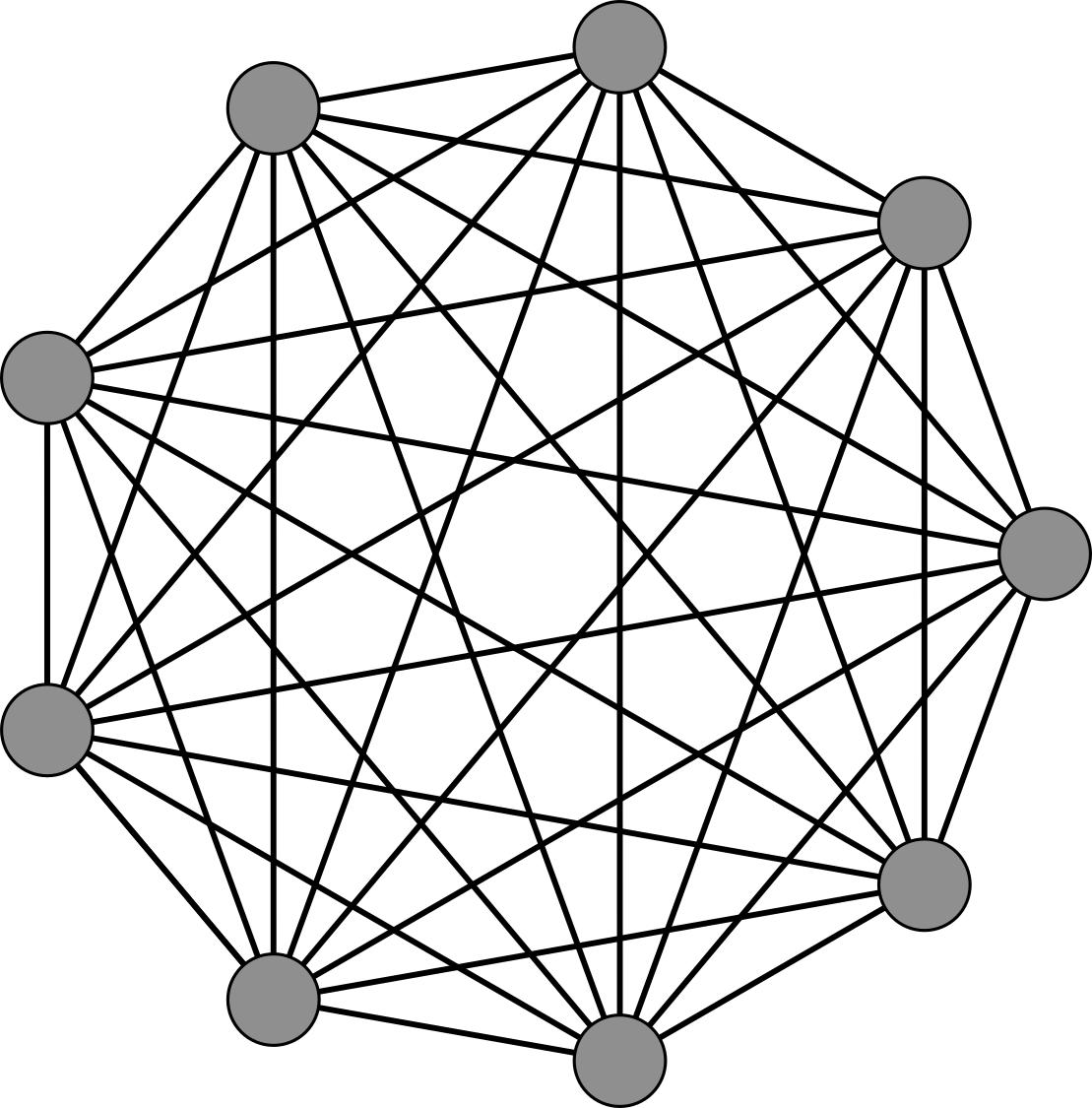}\\
    \caption{Illustration of networks of nine nodes with different topologies: \textbf{(A)} Two-dimensional periodic square lattice (short-range interactions only) \textbf{(B)} Scale-Free Network (both short- and long-range interactions) \textbf{(C)} Clique (all nodes interact with each other).}
    \label{fig:struc}
\end{figure}

However, a scale-free network has connectivity that ranges from nodes that interact short-range only to nodes that may have both short- and long-range interactions (as illustrated in \fig\ \ref{fig:struc} \textbf{(B)}). It can be approximated by a network of interacting spins where the average number of interacting spins on a spin is $\mk = 2\mm$ and the critical temperature $\tc$ as proposed in \sect\ \ref{sec:ordising} as,

\begin{equation}
\tc^{\mathrm{sfnet}} = \frac{2\mm\cc}{\kb}
\label{eq:tcsf}
\end{equation}
Among these three structures, a regular two-dimensional lattice has the lowest connection density. Generally, then, the connectivity structure of a scale-free network is between the topology of a lattice and a clique. Therefore we expect the Ising model of a scale-free network to exhibit a critical temperature between lattice and clique.

Since mean-field calculations approximate the same critical exponent for all the above three topologies, the vital difference is in the critical temperature of $\tc$. For a \banet\ with preferentially attached links, $\mm > 2$, or $3$, the critical temperature is higher than the critical temperature of a two- or three-dimensional lattice with nearest-neighbor interactions. On the other hand, the critical temperature of a \banet\ is less than that of the critical temperature of a clique because $\mm$ should be less than $\frac{(\nn-1)}{2}$ to preserve its scale-free structure. Therefore, $\tc^{\mathrm{lattice}} < \tc^{\mathrm{sfnet}} < \tc^{\mathrm{clique}}$, unless $\mm < 2$.

\begin{figure}[!htb] 
    \centering
    \includegraphics[scale=0.8]{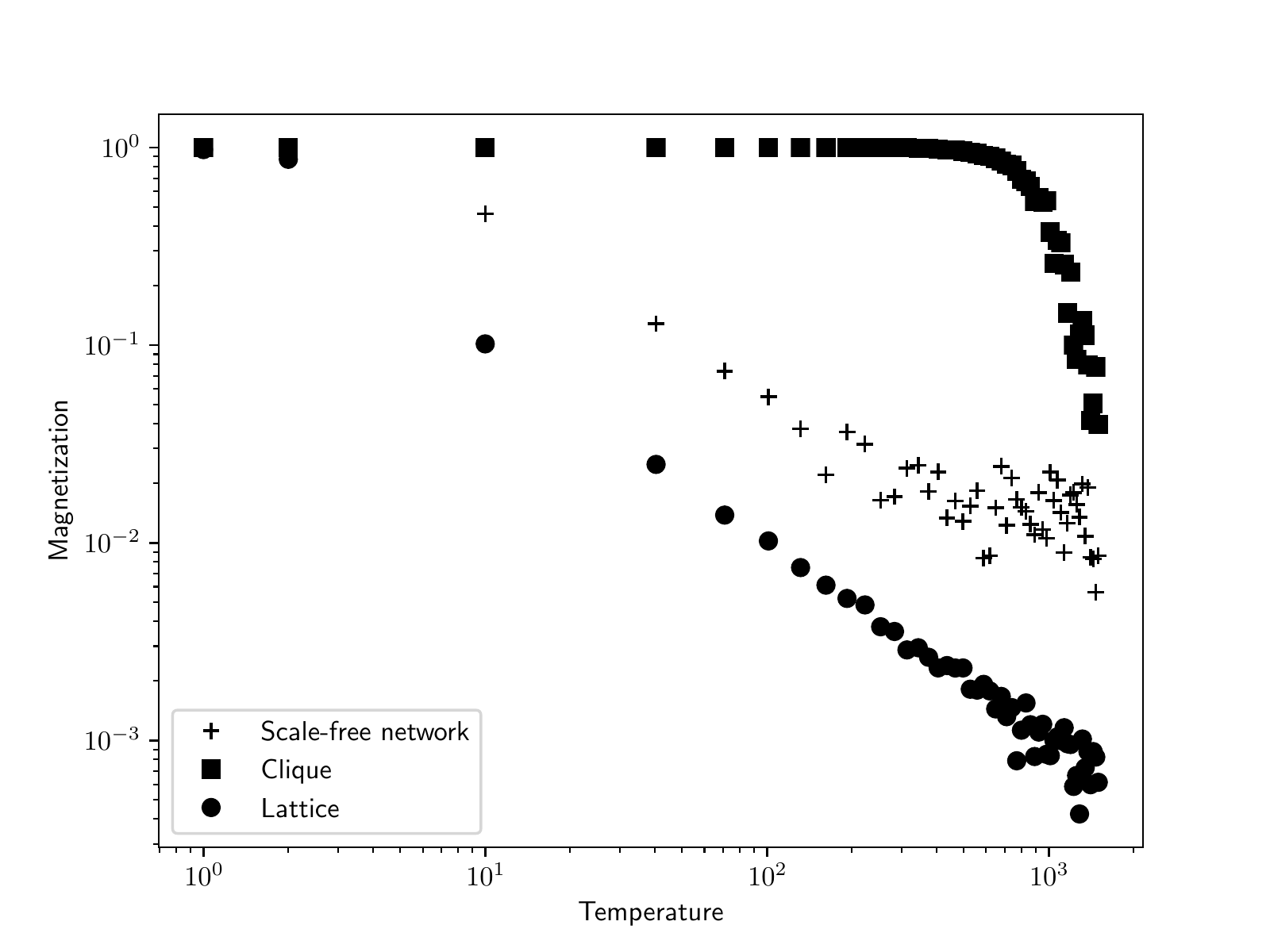}\\
    \includegraphics[scale=0.2]{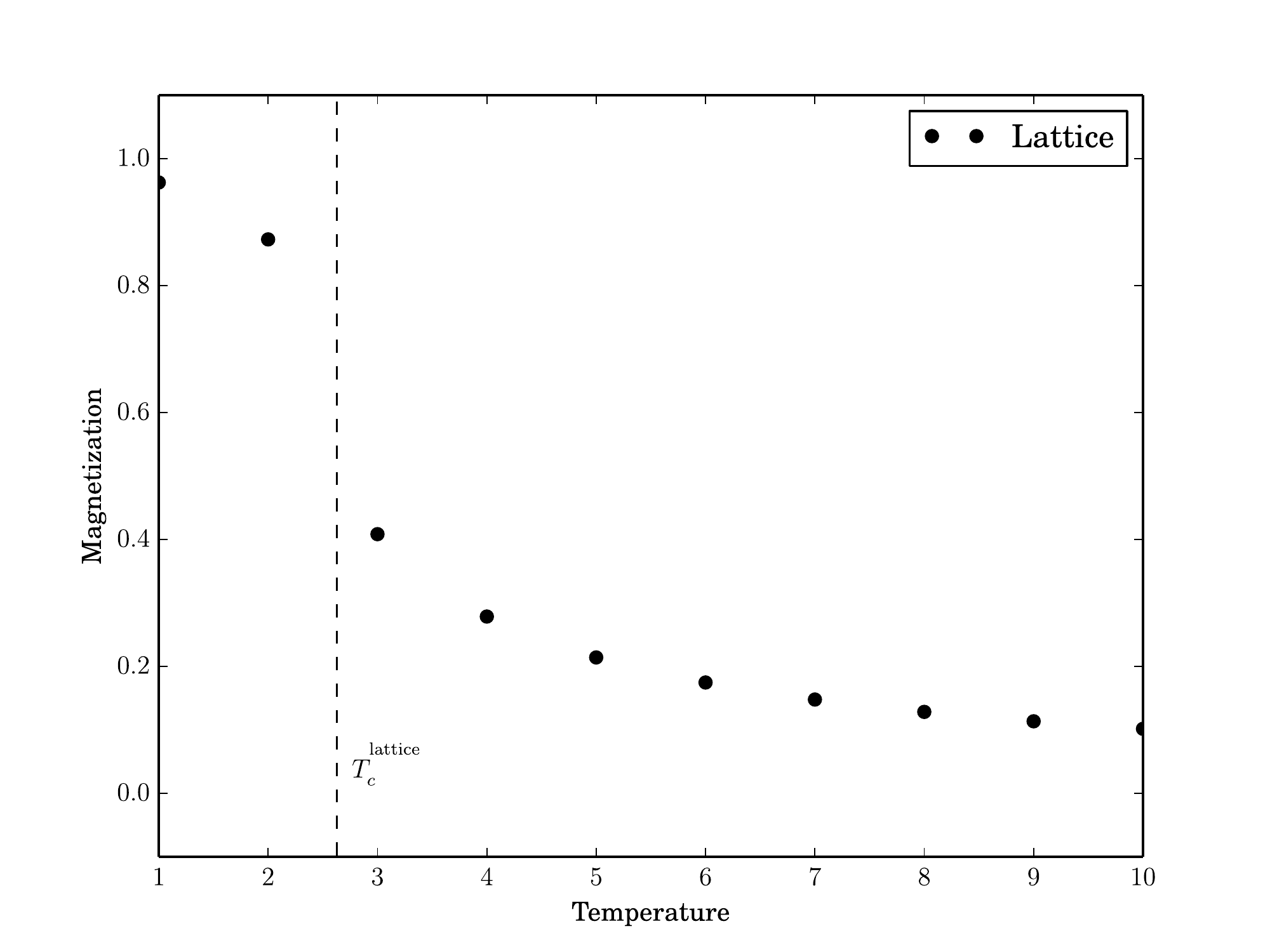}
    \includegraphics[scale=0.2]{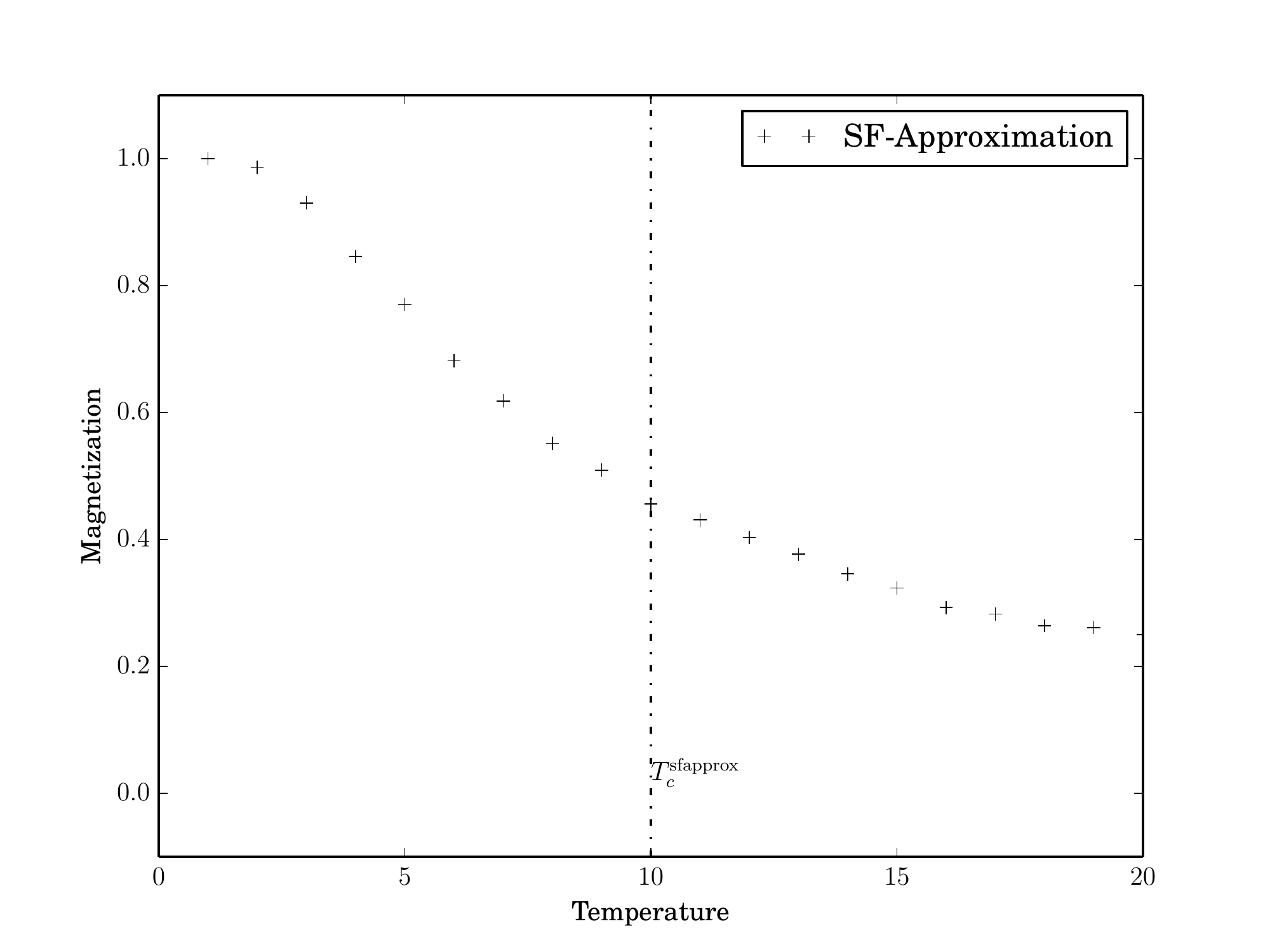}
    \includegraphics[scale=0.2]{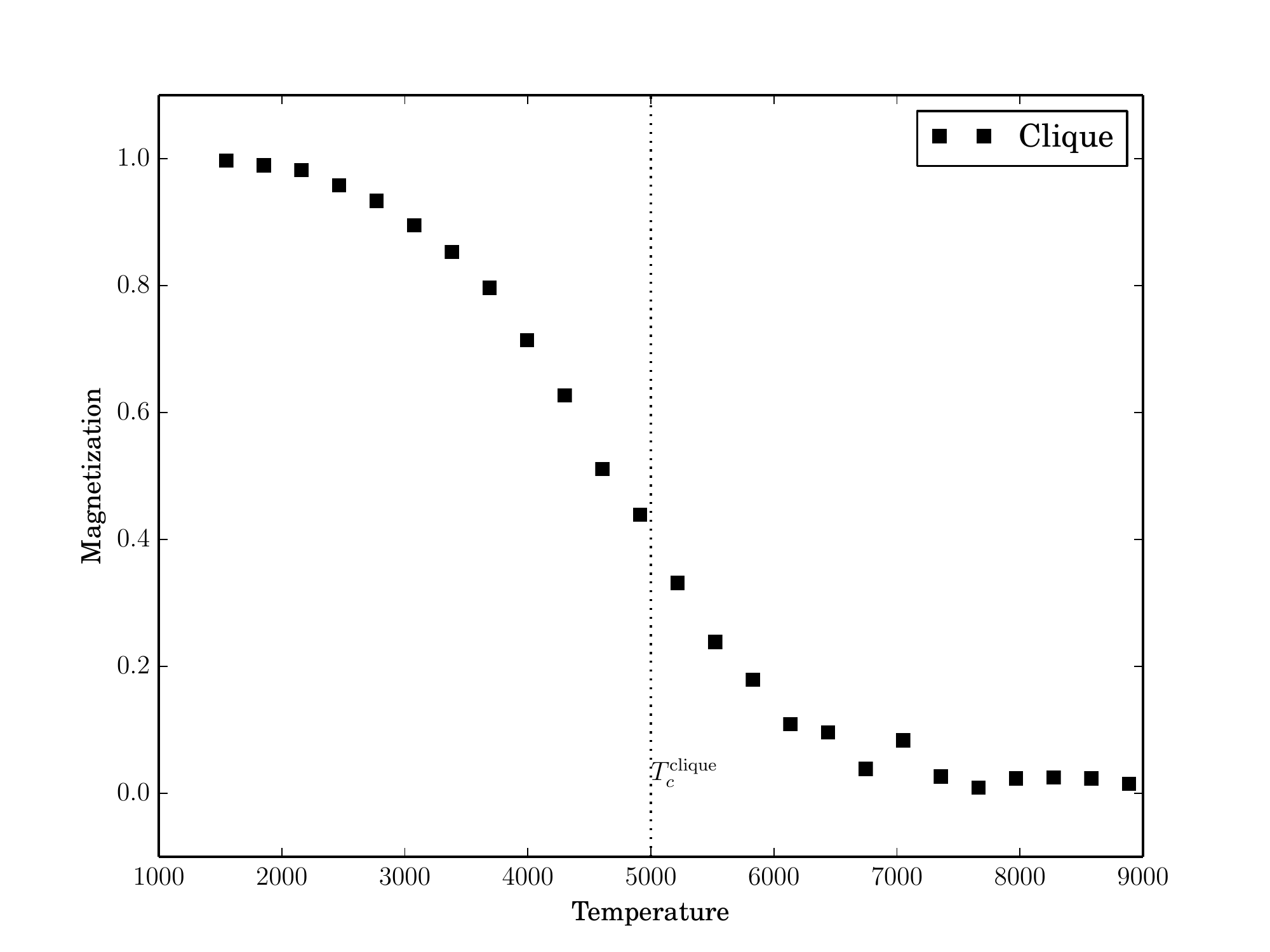}
    \caption{Log-log plot of Monte Carlo simulations of Ising model of \textbf{(A)} two-dimensional Lattice of size $\nn = 1024$ (indicated by circles) \textbf{(B)} \lrim\ of a \ba\ scale-free network of size $N = 1000$ and preferentially attached links, $\mm = 5$ (indicated by plus markers) \textbf{(C)} Clique of size $\nn = 1000$ (indicated by square boxes). Numerical results validate results from mean field calculations (\cf\ \sect\ \ref{sec:ordising}): $\tc^{\mathrm{lattice}} = 2.27$, $\tc^{\mathrm{sfnet}} = 10$, $\tc^{\mathrm{clique}} = 999$. Critical temperature of Ising models of clique and lattice form the upper and lower bounds respectively for critical temperature of a scale-free network.}
    \label{fig:comparison}
\end{figure}

For the same coupling constant $\cc$, the Ising simulations of a scale-free network show a trend between lattice and clique where lattice is the lower bound (undergoes phase transition at low temperatures) and clique is the upper bound (undergoes phase transition at higher temperatures) as can be verified from the Monte Carlo simulations of the Ising models on the three topologies as shown in \fig\ \ref{fig:comparison}. In a regular lattice, the spins are well-connected and hence exhibits spin flips at low temperatures (indicated as circles in \fig\ \ref{fig:comparison}). However, a clique is well-connected; therefore, it requires that the system is heated much longer before spins flip down (indicated as box markers in \fig\ \ref{fig:comparison}). 

Finally in \sect\ \ref{sec:ordising} we showed that the approximation of a Ising model on a \banet\ can be interpreted as a special case of a \lrim\ with a reduced effective coupling $\jeff$ (\eq\ \ref{eq:jeff}). As a corollary we can make the inference that it can be intepreted as a special case of Ising model on a clique as well. Hence the model could be alternatively referred to as the $\mk$-clique model as mentioned in the previous sections. Consider \eq\ \ref{eq:tc2},

\begin{equation}
\setlength{\jot}{10pt}
\begin{split}
\tc &= \frac{2\mm\cc}{\kb}\\
&= \frac{\nn \frac{2\mm\cc}{\nn}}{\kb}\\
\tc &= \frac{\nn \jeff}{\kb} \hspace{0.3cm} \mathrm{where} \hspace{0.3cm} \jeff = \frac{2\mm\cc}{\nn}
\end{split}
\label{eq:cliqsf}
\end{equation}

\Eqs\ \ref{eq:tc2} and \ref{eq:cliqsf} show the relationship between the three topologies in terms of the critical temperature. Phase transition in a complex structure such as the scale-free network is, in a way, fundamentally an alternative formulation of phase transitions in homogeneous structures such as a clique and a lattice.

\cleardoublepage
\section{Comparison of Mean Field Theories for Ising model of \banet} 
\label{sec:mftcomp}

\begin{figure}[!htb] 
\hspace{-2cm} \includegraphics[scale=0.4]{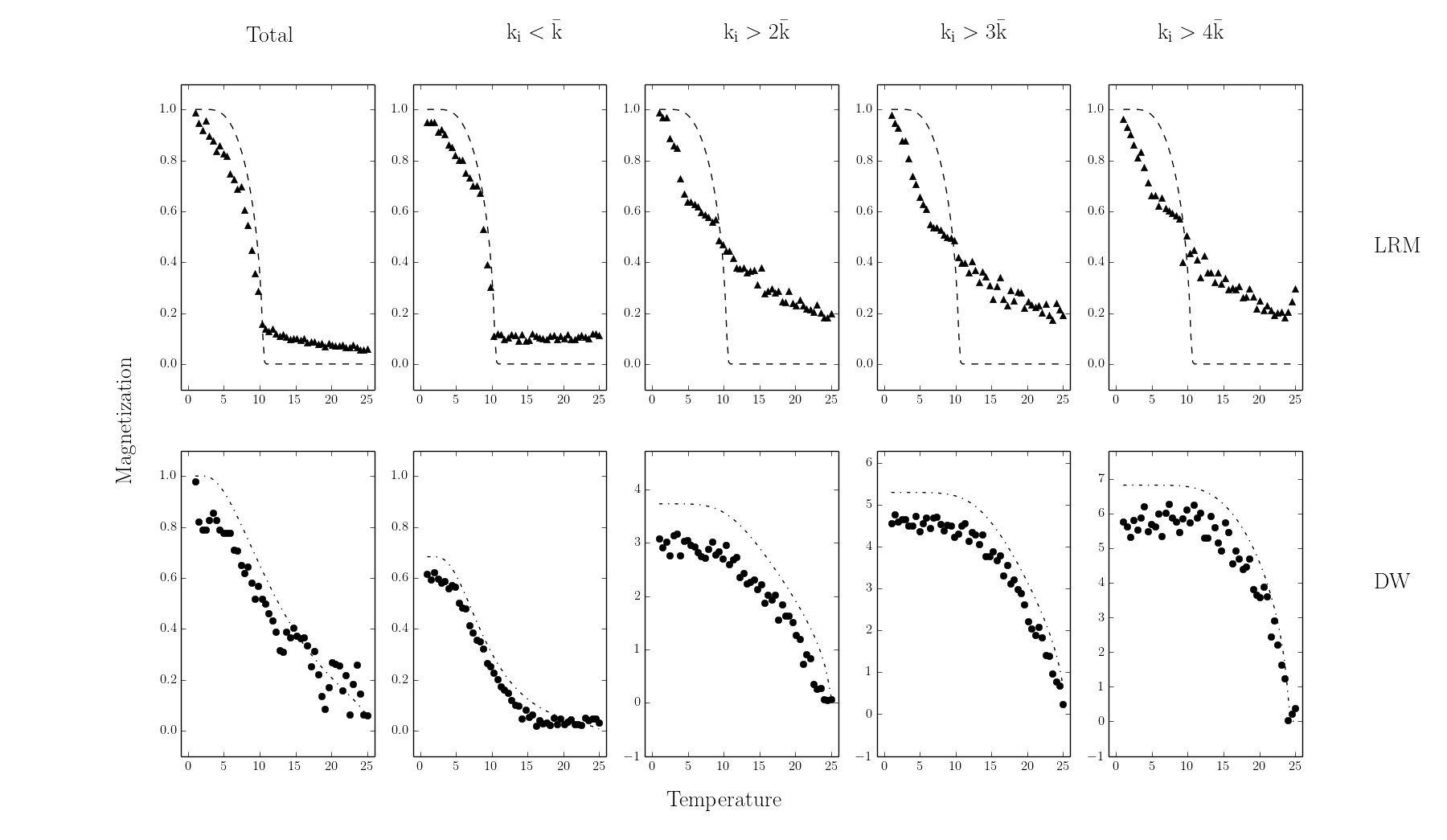}
\caption{Comparison of the \lrim\ (first row, indicated as LRM) and degree-weighted theory (second row, indicated as DW): numerical results are shown by triangle and circle markers respectively. Mean-field approximation is shown by dashed and dotted lines respectively. The first column shows the total magentization of all nodes in the network; second to fifth column shows magnetization of selected nodes of the network of degrees $\kk < \mk$, $\kk > 2\mk$, $\kk > 3\mk$ and $\kk > 4\mk$ respectively where $\mk \approx 10$ for the chosen simulation parameters ($\nn = 5 \times 10^3$, $\mm = 5$, $\cc = 1$). Expected $\tc$ for \lrim\ $= 10$ (\eq\ \ref{eq:tc2}); expected $\tc$ for degree-weighted theory $= 21.29$ \citep{bianconi_2002}.}
\label{fig:fnumtheory}
\end{figure}

The \lrim\ of a \banet\ proposed in \sect\ \ref{sec:ordising} is one of the mean-field formulations for phase transitions occurring in \banet. \lrim\ uses a global averaged mean-field with degree distribution approximated by the mean degree to arrive at the mean-field approximation in \eq\ \ref{eq:mag}. The degree-weighted mean-field approximation, however, uses a node dependent expectation value instead. To compare the two models, we performed Monte Carlo simulations of the \lrim\ of \banet\ (indicated as triangle markers in \fig\ \ref{fig:fnumtheory}) and degree-weighted Ising model of \banet\ (indicated as circle markers in \fig\ \ref{fig:fnumtheory}) for selected nodes of the network to infer how well they fit with their respective mean-field approximations (indicated in dashed and dotted-dashed lines respectively). For a network of size $\nn = 5 \times 10^3$, the system is equilibrated for $2 \times 10^4$ MC steps and thermodynamic variables sampled over $3 \times 10^4$ MC steps. We categorize nodes in the network based on their degree as follows:

\begin{itemize}
\item[-] nodes of all degree ($\kk$);
\item[-]nodes with low degree ($\kk < \mk$);
\item[-]nodes with degree slightly higher than mean degree ($\kk > 2\mk$); 
\item[-]nodes with degree significantly higher than average ($\kk >3\mk$) and
\item[-]nodes with high degree ($\kk > 4\mk$)
\end{itemize}

The total magnetization of \lrim\ and degree-weighted model of \banet\ agree reasonably well with the trend predicted by the mean-field theory (first column of \fig\ \ref{fig:fnumtheory}). The total magnetization of the \lrim\ follows that of nodes with a low degree while the degree-weighted model \enquote{underestimates} magnetization owing to the lower weighting of the nodes (second column of \fig\ \ref{fig:fnumtheory}). However, for nodes with a degree slightly higher than the mean degree or above, \lrim\ exhibits finite magnetization even at very high temperatures and does not agree with mean-field (third, fourth, and fifth columns of \fig\ \ref{fig:fnumtheory} LRM). These high degree nodes that order at high temperatures may form an effective magnetic field for the low-degree nodes that are yet to order, thereby not reaching a paramagnetic state. On the other hand, though the degree-weighted model \enquote{overestimates} magnetization above unity, it accurately predicts the ordering of nodes with a high degree (third, fourth, and fifth columns of \fig\ \ref{fig:fnumtheory} DW). In summary, the \lrim\ describes the magnetization of the majority of sites with mean or smaller than mean degree better, while the degree-weighted theory is better for predicting the onset at higher temperatures. 

\begin{figure}[!htb]
\textbf{(A)}\includegraphics[scale = 0.4]{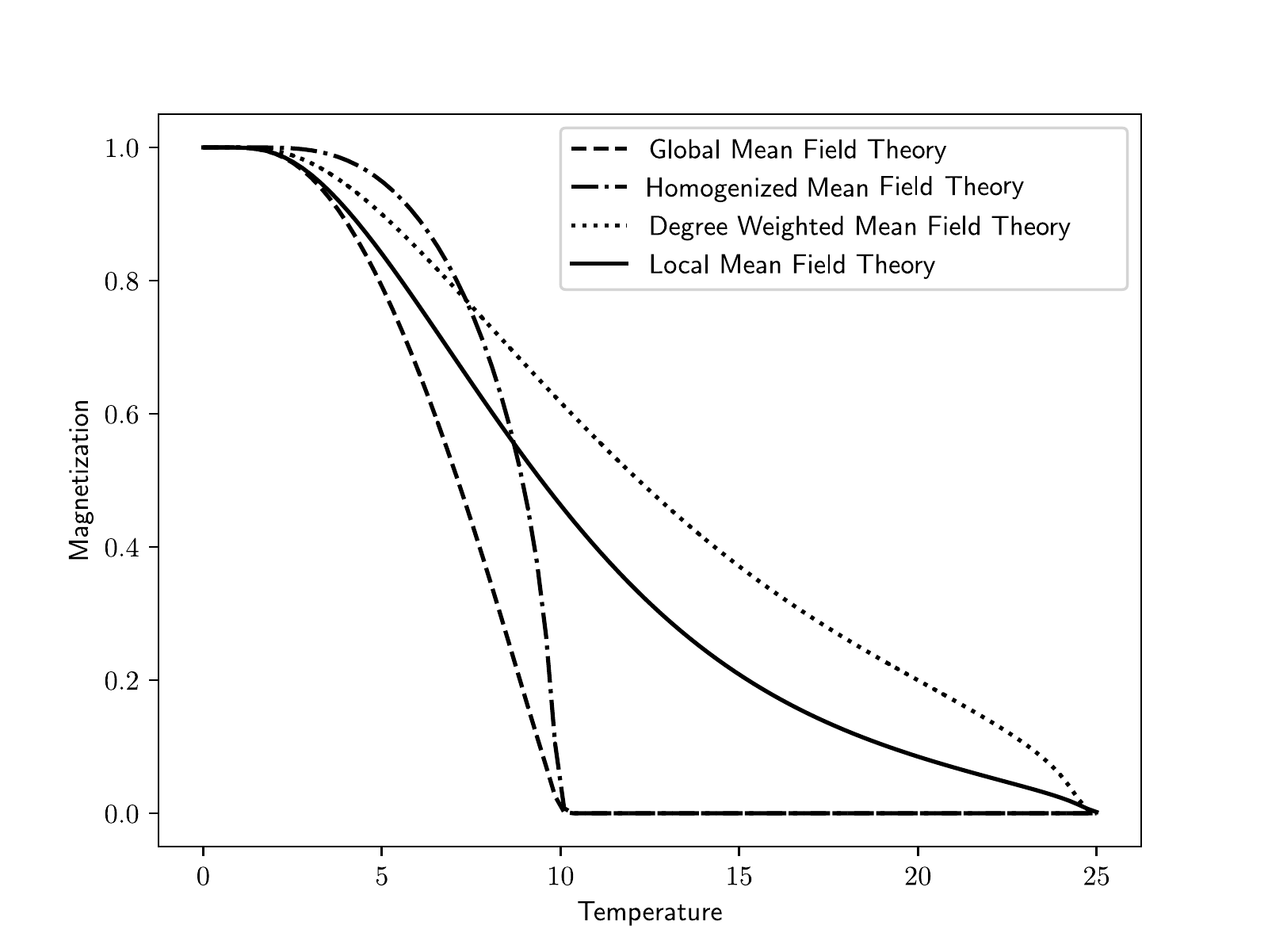}
\textbf{(B)}\includegraphics[scale = 0.4]{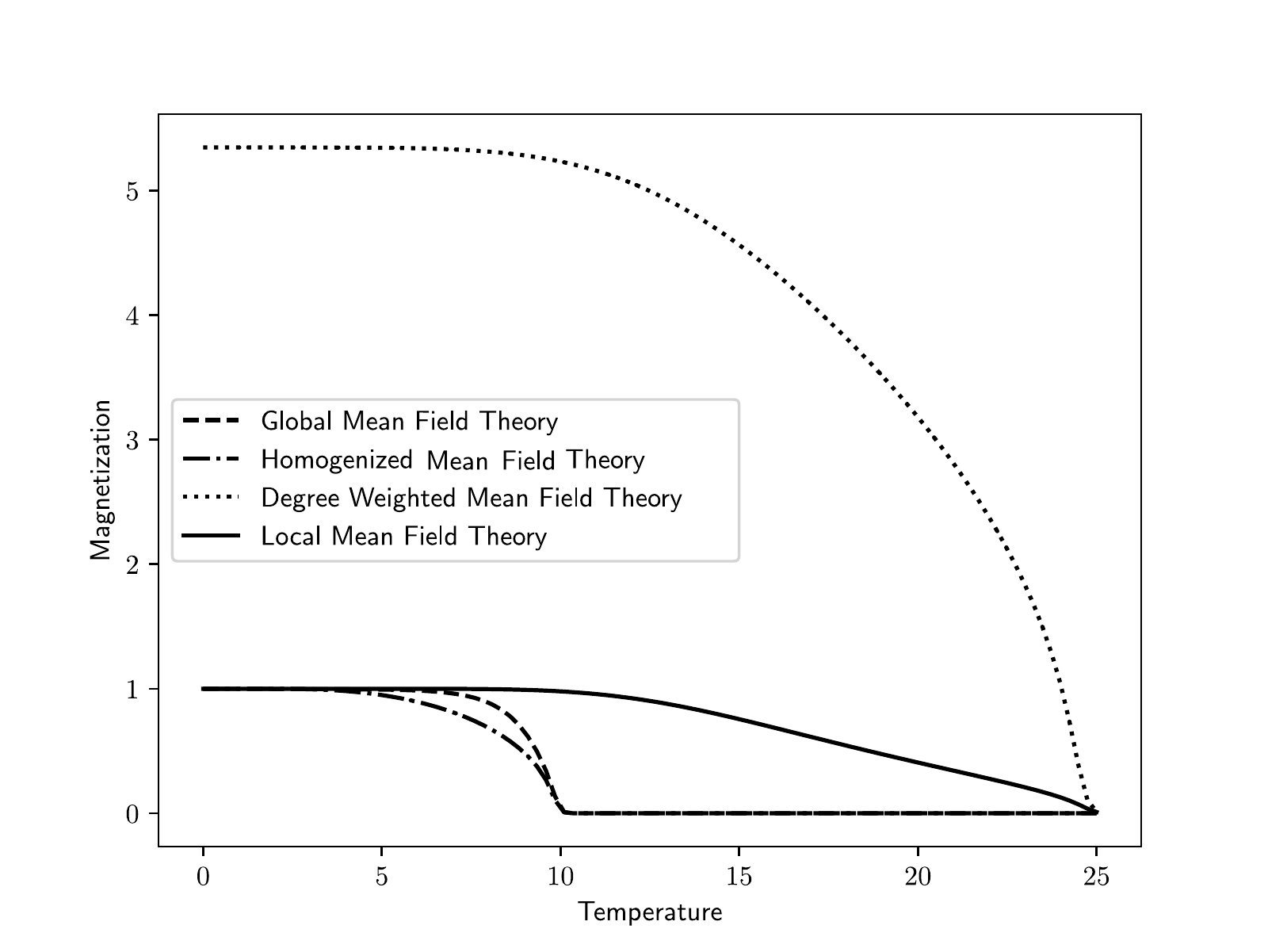}\\
\textbf{(C)}\includegraphics[scale = 0.4]{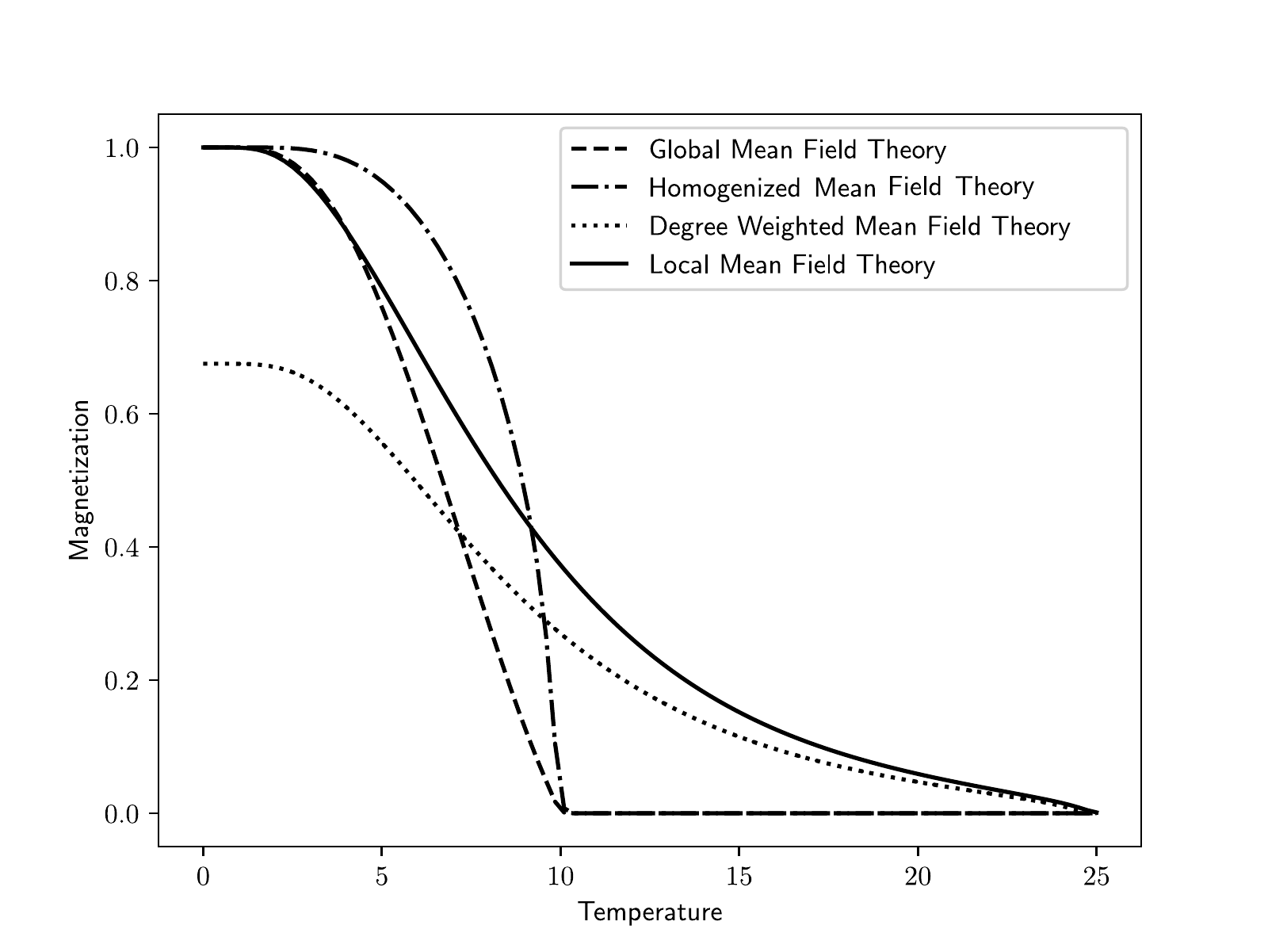}
\caption{Comparison of Mean Field Theories: Figure shows the mean field equations for \textbf{(A)} Total magnetization of all nodes in the network, \textbf{(B)} Net magnetization of nodes with high degree $(\kk_i\geq 3\bar{k})$; (C) Net magnetization of nodes with low degree $(\kk_i < \mk)$. Dashed lines show global mean field theory, dashed lines with dots show homogenized long-range mean field theory, dotted lines show degree weighted or heterogeneous mean field theory and straight lines show local mean field theory in each case. Network parameters: $\nn = 5 \times 10^3$, $\mm = 5$; and coupling constant, $\cc = 1$.}
\label{fig:ftheories}
\end{figure}

For completeness, we also mention here two other possible models - 1. global mean-field, a variant of the \lrim\ with node dependent expectation value where majority of the sites order (\fig\ \ref{fig:ftheories}(A)) and 2. local mean-field, a variant of the degree-weighted model with mean-field being node dependent where  the strongly connected sites order most (\fig\ \ref{fig:ftheories}(B) and summarized in \tab\ \ref{tab:mftcomp1}). Ising model of \banet\ has different energy scales for different mean-field classes.

\begin{table}[!htb]
\begin{tabular}{ |c|c|c| } 
 \hline
 Method & Magnetization & Local spin expectation value\\
 \hline
 Long-range &  $\mg = \frac{1}{\nn}\sum_{i=1}^{\nn}\langle s_i \rangle$ & $\langle s_i \rangle = \tanh(\beta\cc\langle \kk\rangle\mg)$\\ 
 \hline
 Degree-weighted & $\mg = \frac{1}{\nn} \sum_{i=1}^{\nn} \frac{\kk_i}{\langle \kk \rangle} \langle s_i \rangle$ & $\langle s_i \rangle = \tanh(\beta\cc\kk_i\mg)$ \\ 
 \hline
 Global mean-field & $\mg = \frac{1}{\nn}\sum_{i=1}^{\nn}\langle s_i \rangle$ & $\langle s_i \rangle = \tanh(\beta\cc\kk_i\mg)$\\
 \hline
 Local mean-field & $\mg_i = \sum_{j=1}^{\nn}\adj\langle s_i \rangle$ & $\langle s_i \rangle = \tanh(\beta\cc\mg_i)$ \\ 
 \hline
\end{tabular}
\caption{Comparison of Mean Field formulation for phase transitions in occurring in \ba\ network}
\label{tab:mftcomp1}
\end{table}
\clearpage
\section{Conclusions}
\label{sec:concl}

In this work, we have shown that the phase transition behavior of an Ising model on \banet, wherein the order parameter of the system is defined as the ensemble average of all spins of the network, is well-approximated by a \lrim\ wherein each node has neighbors equal to the mean degree of the \banet. Such a model of \banet\ works well at low temperatures and close to critical temperatures ($0 \leq \tp < \tc$); but the approximation appears to be limited at critical temperature ($\tp \geq \tc$) owing to the error arising from nodes of very high degree. The critical temperature for the long-range Ising model scales linearly with \ba\ model parameters and coupling constant of the Ising model. 

This dependence allows us to control the critical behavior of a \ba\ network by changing the model parameters. We infer that the structure (network) and dynamics (the Ising model) are closely interconnected to each other. We can change the control parameter of the Ising model (such as temperature) by changing the network parameters (such as links added), which opens up a new window into the study of the critical behavior of \ba\ networks and real-world networks that have connectivity close to a power-law degree distribution. The \lrim\ describes the magnetization of the majority of the sites with average or smaller than average degrees better compared to the degree-weighted theory, which predicts the onset at higher temperatures better. 

Lattices, scale-free networks, and cliques have traditionally been treated as graphs with different structures that exhibit phase transition behavior that is very far from each other. It has, therefore, not been possible to compare the phase transition behavior between these structures objectively owing to their unique structural connectivity. We have shown that the scale-free network behaves like a clique or a lattice with a reduced effective coupling. This approximation of an Ising model of a scale-free network to a long-range Ising model allows us to make a direct comparison of a scale-free network to simple graphs such as lattices and cliques of the same size. With this, we have shown that the critical temperature of these network structures can be directly mapped from one to another. We infer that the phase transition behavior of regular and complex structures are not very different, and the critical temperature on these structures can be mapped from one topology to another. 

\clearpage
\section{Appendix}
\label{sec:appdx}
\subsection{Approximation of ensemble average of adjacency matrix by network parameters}
\label{subsec:eavg}

\noindent Here we summarize the approach from \cite{bianconi_2002} to reduce mean adjacency matrix over many realization of \banet\ to network parameters. Let us consider a \banet\ of $\nn$ nodes. Starting from a small number of nodes $\nn_{0}$ and links $\mm_{0}$ (where $\nn_{0}, \mm_{0} << \nn$), the network is constructed iteratively by the constant addition of nodes with $\mm$ links. The new links are preferentially attached to well connected nodes in such a way that at time $t_j$, the probability $\pij$ that the new node $j$ is linked to node $i$ with connectivity $k_i(t_j)$ is given by,

\begin{equation}
\pij = \mm \frac{\kk_{i}(t_j)}{\sum_{\alpha = 1}^{j} \kk_{\alpha}}
\label{eq:connprob1}
\end{equation}

\noindent is proportional to the number of links $\kk_i$ at time $t_j$, and number of preferentially attached links $\mm$. The dynamic solution of connectivity at time $t_i$ is,

\begin{equation}
\kk_{i} = \mm \sqrt{\frac{t}{t_i}}
\label{eq:kt}
\end{equation}

\noindent From \eqs \ref{eq:connprob1} and \ref{eq:kt} we have,

\begin{equation}
\pij = \mm \frac{\mm \sqrt{\frac{t}{t_i}}}{\sum_{\alpha = 1}^{j} \kk_{\alpha}(t)}
\label{eq:connprob2}
\end{equation}

If $\nn$ is large we can approximate the total number of edges in the network at time $t_j$, given by the sum $\sum_{\alpha=1}^{j}\kk_{\alpha}$ as, 

\begin{equation}
\sum_{\alpha = 1}^{j} \kk_{\alpha} = \mm_0 + 2\mm t_j \approx 2\mm t_j
\label{eq:kalpha} 
\end{equation}

\noindent because $\mm_{0} << \nn$. The factor $2$ comes from the fact that as we create a link which connects two nodes, the number of links of each of them increases by $1$. Substituting \eq\ \ref{eq:kalpha} in \ref{eq:connprob2},

\begin{equation}
\begin{split}
\pij &= \frac{\mm^2 \sqrt{\frac{t_j}{t_1}}}{2\mm t_j}\\
&= \frac{\mm}{2} \frac{1}{\sqrt{t_it_j}}\\
\label{eq:connprob3}
\end{split}
\end{equation} 

\noindent The adjacency elements of the network $\adj$ are equal to $1$ if there is a link between node $i$ and $j$ and $0$ otherwise. Consequently the mean over many copies of a \banet\,

\begin{equation}
\langle \adj \rangle = \pij = \frac{\mm}{2}\frac{1}{\sqrt{t_it_j}}
\label{eq:adjmean}
\end{equation}

\noindent From \eq\ \ref{eq:kt} we can re-write for $t = \nn$ steps,

\begin{equation}
\begin{split}
\kk_{i}(t) &= \mm \sqrt{\frac{t}{t_i}}\\
\kk_{i}(\nn) &= \mm \sqrt{\frac{\nn}{t_i}}\\
t_i &= \frac{\mm^2 \nn}{\kk_{i}^{2}}
\label{eq:ti}
\end{split}
\end{equation}

\noindent and similarly,

\begin{equation}
t_j = \frac{\mm^2\nn}{\kk_{j}^{2}}
\label{eq:tj}
\end{equation}

\noindent From \eqs\ \ref{eq:ti} and \ref{eq:tj}, 

\begin{equation}
\begin{split}
\langle \adj \rangle &= \frac{\mm}{2} \frac{1}{\sqrt{\frac{\mm^2\nn}{\kk_{i}^{2}}}\sqrt{\frac{\mm^2\nn}{\kk_{j}^{2}}}}\\
&= \frac{1}{2\mm\nn}\kk_{i}\kk_{j}\\
\label{eq:adjnp1}
\end{split}
\end{equation}

\noindent The average of the adjacency matrix over many realizations can be approximated by the network parameters as,

\begin{equation}
\langle \adj \rangle = \frac{1}{2\mm\nn}\kk_{i}\kk_{j}
\label{eq:adjnp2}
\end{equation}

\noindent The mean degree of the network $\mk$ can be approximated from \eq\ \ref{eq:kalpha} as,

\begin{equation}
\begin{split}
\mk &= \frac{1}{\nn}\sum_{i=1}^{\nn}\kk_{i}\\
&= \frac{1}{\nn} 2\mm\nn \\
\mk &\approx 2\mm\\
\label{eq:meank}
\end{split}
\end{equation}
\subsection{Degree Distribution Deviation}
\label{subsec:degdev}

Consider the sum of degree distribution deviation summed over $i$ (as in \eq\ \ref{eq:t0}),

\begin{equation}
\begin{split}
\sum_{i=1}^{\nn} \delta \kk_i &= \sum_{i=1}^{\nn} (\kk_i - \mk)\\
&= \sum_{i=1}^{\nn}\kk_i - \sum_{i=1}^{\nn}\mk = 0\\
\end{split}
\end{equation}

since $\mk = \frac{1}{\nn}\sum_{i=1}^{\nn}\kk_i$. Hence the average over deviation of degree distribution from mean degree is zero.

\section{Funding}
This work was supported by the Exploratory Research Space (ERS) Seed Fund 2017 in Computational Life Sciences (CLS001). All simulations were performed using the RWTH Compute Cluster under general use category; priority category allocated to AICES and JRC users; and with specific computing resources granted by RWTH Aachen University under project rwth0348. The authors gratefully acknowledge the generous support of the aforementioned funding and computing resources.
\section{Acknowledgments}

JK thanks Richard Polzin for his help with the network illustration in this paper; and Ajay Mandyam Rangarajan for the many useful discussions.
\include{biblio}
\end{document}

%% file: preamble.tex
\usepackage{amsmath}  
\usepackage{amsfonts} 
\usepackage{graphicx}
\usepackage{xcolor}
\usepackage{rotating}
\usepackage{booktabs}
\usepackage{csquotes}
\usepackage[colorlinks=true,allcolors=blue]{hyperref}
\usepackage{tikz} 
\usetikzlibrary{spy}

\usepackage{fancyhdr}


%% file: macros.tex
\newcommand{\kk}{k}						
\newcommand{\hm}{H}						
\newcommand{\cc}{J}						
\newcommand{\adj}{\mathrm{A_{ij}}}		
\newcommand{\hh}{h}						
\newcommand{\si}{s_i}					
\newcommand{\sj}{s_j}					
\newcommand{\nn}{N}						
\newcommand{\mm}{m}						
\newcommand{\mk}{\bar{k}}				
\newcommand{\dk}{\delta{k}}				
\newcommand{\tc}{T_c}					
\newcommand{\kb}{k_B}					
\newcommand{\jeff}{J_{\mathrm{eff}}}	
\newcommand{\mg}{M}						
\newcommand{\zz}{z}						
\newcommand{\pij}{p_{ij}}				
\newcommand{\cm}{J_{\mathrm{ij}}}		
\newcommand{\tp}{T}						
\newcommand{\hmf}{H_{MF}}				
\newcommand{\bt}{\beta}					
\newcommand{\lrim}{long-range Ising model} 
\newcommand*{\banet}{Barab\'{a}si-Albert Network}
\newcommand*{\ba}{Barab\'{a}si-Albert}

\newcommand*{\ie}{i.e.}
\newcommand*{\cf}{cf.}
\newcommand*{\fig }{Fig.} 
\newcommand*{\figs }{Figures} 
\newcommand*{\sect }{Sec.} 
\newcommand*{\ssec }{Subsec.} 
\newcommand*{\sssec }{Subsubsec.} 
\newcommand*{\eq }{Eq.}
\newcommand*{\Eqs }{Eqs.}
\newcommand*{\tab }{Table}
\newcommand*{\eqs }{Eqs.}
\newcommand*{\app }{Appendix}